\documentclass[11pt, letterpaper]{article}

\pdfoutput=1

\usepackage{amsmath}
\usepackage{amsfonts}
\usepackage{amssymb}
\usepackage[colorlinks,citecolor=blue,urlcolor=blue,bookmarks=false,hypertexnames=true]{hyperref}
\usepackage{cite}

\usepackage{graphicx, physics, subcaption}
\usepackage{verbatim }

\numberwithin{equation}{section}

\usepackage{color}

\usepackage{amsmath, amssymb, graphics, epsfig, graphicx}
\usepackage{epsf}
\usepackage{epstopdf}
\usepackage {amssymb}
\newcommand{\nc}{\newcommand}
\nc{\ba}{\begin{eqnarray}}
\nc{\ea}{\end{eqnarray}}

\def\bfk{{\bf k}}
\def\bfq{{\bf q}}
\def\bfx{{\bf x}}

\nc{\GeV}{ \mathrm{GeV}}
\nc{\eV}{ \mathrm{eV}}
\nc{\Hm}{  H_{(m)} }
\nc{\HF}{  H_{F} }

\begin{document}

\vspace{5mm}
\vspace{0.5cm}
\begin{center}

\def\thefootnote{\fnsymbol{footnote}}

{\bf\Large Cosmological constant problem on the horizon}
\\[0.5cm]

{ 
Hassan Firouzjahi$^{1}\footnote{firouz@ipm.ir},$
}
\\[0.5cm]

{\small \textit{$^{1}$ School of Astronomy, Institute for Research in Fundamental Sciences (IPM) \\ P.~O.~Box 19395-5531, Tehran, Iran
}}\\

\end{center}

\vspace{.8cm}

\hrule \vspace{0.3cm}


\begin{abstract}

We revisit the quantum cosmological constant problem and highlight the important roles played by the dS horizon of  zero point energy. We argue that fields which are light enough to have dS horizon of zero point energy 
comparable to the FLRW Hubble radius are the main contributors to dark energy. 
On the other hand, the zero point energy of heavy fields develop nonlinearities on sub-Hubble scales and can not contribute to dark energy. We speculate that our 
proposal may provide a  resolution for both the old and new cosmological constant problems by noting that there exists a field, the (lightest) neutrino, which happens to have a mass comparable to the present background photon temperature. 
The proposal predicts multiple transient periods of dark energy in early and late expansion history  of the universe yielding to a higher value of the current Hubble expansion rate which  can resolve the $H_0$ tension problem.

\end{abstract}
\vspace{0.5cm} \hrule
\def\thefootnote{\arabic{footnote}}
\setcounter{footnote}{0}
\newpage

\section{Introduction}
\label{sec-intro}

The $\Lambda$CDM model  has emerged as a successful phenomenological model, explaining the dynamics of the evolution of the cosmos with a handful of free parameters \cite{Weinberg:2008zzc, Planck:2018vyg}. 
Among the key unknown ingredients are dark matter and dark energy. There are hopes that the former may be addressed one way or another by new physics beyond the Standard Model (SM) of particle physics while  the latter has proven more elusive. 

There have been numerous attempts to address the nature of dark energy,
for a review see \cite{Sahni:1999gb, Copeland:2006wr, Peebles:2002gy} and references therein. One simple and natural possibility is that the dark energy is simply a cosmological constant term, like what Einstein initially introduced 
to construct a seemingly static universe. As is well known, the cosmological constant is associated with big problems in theoretical physics, for a classic review see \cite{Weinberg:1988cp}. As we shall review  below, it is expected  that the cosmological constant receives contributions from the zero point energy of all fields in SM. It is usually argued that the zero point energy density is at the order $M^4$ in which $M$ is the UV cutoff of theory. The old cosmological constant problem is why it is not so large as expected from the typical energy scale of the high energy physics, like $\mathrm {(TeV)^4}$ or $M_P^4$ in which $M_P \sim 10^{18} \mathrm {GeV}$ is the reduced Planck mass related to the Newton constant $G$ via  $M_P^2=1/8 \pi G$.
  The new cosmological constant problem is why it is comparable to the energy density of matter at the current epoch in cosmic history, entering to the dynamics of the expansion of the universe at redshift around $z_\Lambda \sim 0.3$.

A number of approaches were proposed to address the cosmological constant problem(s). One approach is to employ the fundamental symmetries, like supersymmetry. In supersymmetric theories there is a one to one relation between the number of bosons and fermions which keep the value of cosmological constant zero. Of course, the Nature is not seen to be supersymmetric on energy scale below  $\mathrm {TeV}$  so the supersymmetry can not explain the smallness of dark energy. The other approach is based on ideas like quintessence \cite{Zlatev:1998tr, Armendariz-Picon:2000nqq}
in which the dark energy is dynamically evolving to is attractor value which is zero. However, these kind of models can not address the severe fine-tunings involved as one has to tune the model parameters so carefully to obtain a tiny value of dark energy at the current epoch. Another class of solutions are based on the self-tuning idea in the context of extra dimensions \cite{Arkani-Hamed:2000hpr, Kachru:2000hf, Csaki:2000dm, Cline:2000ky, Cline:2001yt}. 
These solutions were not convincing either as the solutions either run into singularities or require fine-tuning of model parameters. 
A solution of different kind has been proposed based on anthropic principle in which the cosmological constant should choose the range of values to allow the gravitationally bound objects like galaxies  to form to host star formation required for  the existence of intelligent observer \cite{Weinberg:1988cp, Weinberg:1987dv}. 

In this paper we revisit the quantum cosmological constant problem. We emphasize on  the important roles played by the scale of dS horizon  associated  with the zero point energy. In particular we compare the scale of its Hubble radius with the Hubble radius of the FLRW universe and argue that the conventional treatment of the quantum cosmological constant problem misses this important effect. 

The rest of the paper is organized as follows. In Section \ref{cc-review}
we review the cosmological constant problem and gather some basic formulas. In Section \ref{horizon-sec} we highlight the important roles that the horizon scale of the patch of zero point energy play while in Section \ref{variance-sec} we 
show why, unlike in conventional treatment,  the heavy fields can not contribute to the cosmological constant. In Section \ref{zero-dS} we calculate the zero point energy in dS background while in Section  \ref{implications-sec} we present various cosmological implications of our proposal followed by the Summary and Discussions  in Section \ref{sum-sec}.

\section{The Cosmological Constant Problem} 
\label{cc-review}

In this section we briefly review the cosmological constant problem.
For more extensive review see Refs. \cite{Weinberg:1988cp} and \cite{Martin:2012bt}. 

Numerous observations confirm the detection of dark energy at a level comparable to the matter energy density today 
\cite{SupernovaCosmologyProject:1998vns, SupernovaSearchTeam:1998fmf, Planck:2018vyg}.  Therefore,  the cosmological constant, if not exactly zero by some symmetry considerations,  is as small as $\rho_v \sim (10^{-3} \eV)^4$. 
Now the big  trouble with the cosmological constant is what mechanism keeps it so small  to be consistent with cosmic evolution. 

In its simplest form, the cosmological constant is associated with the problem that the zero point energy of the fields has quartic divergence with the UV energy scale.  To see it more specifically, consider a real scalar field $\phi$ with the mass $m$ described by the following simple Lagrangian in flat spacetime,
\ba
\label{action-scalar}
S= \int d^4 x \Big[ -\frac{1}{2} \partial_\mu  \phi \partial^\mu \phi  - \frac{m^2}{2} \phi^2
\Big] \, .
\ea
The energy momentum tensor $T^{\mu \nu}$ is given by
\ba
\label{Tmunu}
T_{\mu \nu} = \partial_\mu \phi \partial_\nu \phi
 - \eta_{\mu \nu} \Big[ \frac{1}{2} g^{\alpha \beta} 
 \partial_\alpha \phi \partial_\beta \phi + \frac{m^2}{2} \phi^2
 \Big]  \, .
\ea
In particular, the energy density $\rho$ is given by 
\ba
\label{rho}
\rho= T_{00} = {\cal H} = \frac{\dot \phi^2}{2} + \frac{1}{2} \delta^{ij} \partial_i \phi \partial_j \phi + \frac{m^2}{2} \phi^2 \, ,
\ea
in which ${\cal H}$ is the Hamiltonian density.

We expand the quantum field  in terms of the creation and the annihilation operators in Fourier space, 
\ba
\label{Fourier}
\phi(\bfx) = \frac{1}{( 2 \pi)^{3/2}}\int \frac{d^3 \bfk}{\sqrt{2\omega(k)}}  \Big[ e^{i k\cdot x} a_\bfk
+ e^{-i k\cdot x} a_\bfk^\dagger\Big] \, ,
\ea 
in which $k^\mu = (\omega(k), \bfk)$, 
$\omega(k) \equiv \sqrt{\bfk^2 + m^2} $ and $[ a_\bfk, a_\bfq^\dagger] =  \delta^3 (\bfk -\bfq)$. 

The effective energy density contributing to the Einstein field equation is   
$ \langle 0| \rho |0\rangle  \equiv \langle \rho \rangle $ in which $ \langle0| X |0\rangle$ means the quantum expectation value of the quantum operator $X$ with respect to the vacuum. Using the form of the 
operator $\rho$ from Eq. (\ref{rho}) we obtain 
\ba
\label{rhov}
\langle \rho \rangle  = \frac{1}{2}\int \frac{ d^3 \bfk}{( 2 \pi)^3} \omega(k)
  \, .
\ea

Performing the same calculation for the pressure $p$, we obtain \cite{Martin:2012bt}
\ba
\label{p}
\langle p \rangle= \Big \langle \frac{1}{3} \perp^{\mu \nu} T_{\mu \nu} \Big \rangle 
=  \frac{1}{6}\int \frac{ d^3 \bfk}{( 2 \pi)^3} \frac{\bfk^2}{\omega(k)} \, ,
\ea
in which $\perp^{\mu \nu}  \equiv \eta^{\mu \nu} + u^\mu u^\nu$ is a projection operator  and $u^\mu = (1, \bf0)$ is the comoving four-velocity. 

To perform the above integrals, one usually imposes a hard UV momentum cutoff
$k\leq M$ and the integral in Eq. (\ref{rhov}) is estimated as 
\ba
\label{rho-hard}
\langle \rho \rangle \simeq \frac{M^4}{16 \pi^2}\, .
\ea
This is the usual representation of the old cosmological constant problem. If one assumes $M \sim M_P$ then the above estimation yields  a contribution to  $\langle \rho_v \rangle $
larger than the observed value by some factor of $10^{120}$. Of course, if one chooses a smaller UV energy scale, say the scale of SM particle physics $M \sim 10^2 \mathrm {GeV}$, then the discrepancy becomes less severe but still the theoretical estimation is larger by some 
$10^{56}$ orders of magnitude from the observed value. 

The above analysis can be repeated for other fields, bosonic or fermionic, with the important distinction that for fermionic fields one obtains a negative contribution in $\langle \rho \rangle $. This is  because the fermionic fields, unlike the bosonic fields,  anti commute with each other.  

However, as pointed out in \cite{Akhmedov:2002ts, Ossola:2003ku, Koksma:2011cq, Visser:2016mtr, Martin:2012bt}, the hard momentum cutoff implemented above is problematic as it does not respect the Lorentz invariance of the setup.
In other words the hard cutoff only respects the $O(3)$ invariance over the three-dimensional momentum space while breaking the four-dimensional  $O(1, 3)$ Lorentz invariance. To see the fatal problem with this naive hard momentum cutoff regularization, note that to leading order  we obtain 
$\langle p \rangle \simeq M^4/(3\times 16 \pi^2)$ which means that $\langle p \rangle/ \langle \rho \rangle \simeq 1/3$. On the other hand, the Lorentz invariance of the vacuum requires that 
\ba
\langle T_{\mu \nu} \rangle = - \rho_{\mathrm{vac}}\,  \eta_{\mu \nu} \, ,
\ea 
which immediately enforces $\langle p \rangle = - \langle \rho \rangle$. 
To bypass this problem one has to employ a regularization scheme which respects the underlying Lorentz invariance. A good candidate is the dimensional regularization approach which keeps the four dimensional Lorentz invariance intact.
 
Performing the integral using dimensional regularization
approach (or any Lorentz invariant scheme) for the massive real scalar field one obtains  \cite{Akhmedov:2002ts, Ossola:2003ku, Koksma:2011cq, Visser:2016mtr, Martin:2012bt}  
\ba
\label{cc-formula1}
\langle \rho \rangle = -\langle p \rangle=   \frac{m^4}{64 \pi^2} \ln( \frac{m^2}{\mu^2})  \, ,
\ea
in which $\mu$ is the renormalization scale. 

There are a few comments in order concerning Eq. (\ref{cc-formula1}).
First, the requirement of Lorentz invariance $\langle p \rangle = - \langle \rho \rangle$ is explicit. Second, the massless fields such as gravitons, 
photons or gluons do not contribute to the energy density of vacuum. This may be understood by noting that for massless particles the equation of state is
simply $p =\rho/3$ so the requirement of Lorentz invariance $\langle p \rangle = - \langle \rho \rangle$ can be met only if $ \langle \rho \rangle=0$. Third, depending on the renormalization scale, the vacuum energy density can be either positive (dS spacetime) or negative (AdS spacetime). Again, this is unlike the conclusion  based on hard momentum cutoff prescription in which the 
bosonic (fermionic) fields always yield positive (negative) contributions to 
$\langle \rho \rangle$. Finally, there is no quartic dependence on the cutoff scale. This can reduce the naive fine-tuning of order $10^{-120}$ by many orders of magnitudes. As a rough estimate, for the heavy SM fields with mass at the order $10^2 \, \GeV$
we obtain $| \langle \rho \rangle | \sim 10^{44}\,  \eV^4$ which is much smaller than the naive estimation $ \langle \rho \rangle \sim M_P^4$. However, it is still vastly larger than the observed value of the vacuum energy density. In addition, note that the contribution from the  (lightest) neutrino with $m_\nu \sim 10^{-2} \eV$ is $ \langle \rho \rangle \sim m_\nu^4 \sim  (10^{-2} \eV)^4$ which is at the same order as the observed value of dark energy. 
As we shall see this is not an accident and will be part of our solution  for the cosmological constant  problem!

Combining the contributions from all fields, bosonic and fermionic, the total contribution in vacuum zero point energy is obtained to be \cite{Martin:2012bt, Koksma:2011cq}
\ba
\label{cc-formula2}
\langle \rho \rangle = \sum_i n_i \frac{m_i^4}{64 \pi^2} \ln( \frac{m_i^2}{\mu^2}) \, ,
\ea
in which $n_i$ represents the degree of freedom (polarizations) of each field. For example, for a real scalar field $n=1$, for a massive vector field $n=3$   while for a Dirac fermion $n=-4$.  Note that while a fermionic field has an opposite sign contribution  in $\langle \rho \rangle $ compared to a bosonic field, but it is the combination $n_i \ln( \frac{m_i^2}{\mu^2})$ which determines the sign of the contribution of the corresponding field in $\langle \rho \rangle $. For example, if we take $\mu $ to be at the order  of electroweak symmetry breaking scale then all fermionic (bosnic) fields in SM spectrum contribute positively (negatively)  in $\langle \rho \rangle $.

As discussed above, we comment that while a hard  momentum cutoff breaks the underlying Lorentz invariance and yield to a number of problems, such as quartic and quadratic divergences in $\langle \rho \rangle $, but with careful considerations one can still employ a hard momentum cutoff. For this purpose, one has to implement non-covariant counter terms in the Lagrangian to cancel the corresponding power law divergences  in $\langle \rho \rangle $.  More specifically, expanding $\langle \rho \rangle $ in terms of the momentum cutoff scale $M$, the leading divergences are $M^4$ and $M^2$ followed by the logarithmic contribution as in
Eq. (\ref{cc-formula1}). The quartic divergence has the equation of state of radiation while the quadratic divergence has that of a spatial curvature. As emphasized in 
\cite{Martin:2012bt} it is only the logarithmic contribution which has the equation of state of the vacuum which is what captured by the dimensional regularization scheme.

\section{The Question of dS Horizon }
\label{horizon-sec}

To calculate the vacuum energy density we have assumed a flat background yielding to  Eq. (\ref{cc-formula1}).  This seems reasonable as the energy density is a local quantity which is not sensitive to the large scale properties of the cosmological background. Indeed, the combination of Lorentz invariance and the equivalence principle guarantee that the flat spacetime  approximation to calculate the vacuum energy density as given in (\ref{cc-formula1}) is valid. However, there is a subtle issue that was not  taken into account when applying the estimated zero point energy  to cosmology. In the presence of gravity, i.e. when the Newton constant $G$ is turned on, associated to each positive vacuum energy there is a horizon radius which controls the causal properties of the corresponding dS spacetime. 
However, in the usual treatment of cosmological constant problem,  it is assumed that the  {\it entire} observable universe  with the current Hubble radius $H_0^{-1}$ is {\it simultaneously}  endowed with a vacuum energy density $\rho_v$ given by (\ref{cc-formula1}). This is equivalent to assuming  that  the quantum zero point energy fills the entire observable universe in a single patch of size $H_0^{-1}$. This is too much to ask!
 
The correct way of thinking about this problem is to look at the Hubble radius associated to Eq. (\ref{cc-formula2}) for each field. Let us denote  $\rho_v(m)$ and $H_{(m)}$
respectively as the vacuum energy density and the  Hubble rate of dS spacetime created from the zero point energy of each field with the mass $m$. Then we  have $3 M_P^2 \Hm^2 = \rho_v(m) \sim m^4  $, yielding 
\ba
\label{Hm}
\Hm \simeq \frac{m^2}{M_P} \, .
\ea
This is one important relation to keep in mind. For example, for electron field 
with $m_e \sim \rm{MeV}$, we obtain $H_{(m_e)} \sim 10^{-15} \eV$ yielding  the Hubble radius $H_{(m_e)}^{-1} \sim 10^9 m$. 
Logically, this is the largest dS patch which the zero point energy associated with the electron field can cover coherently. On the other hand the current Hubble radius 
is $H_0^{-1} \sim \mathrm{Gpc} \sim 10^{26} m$.  From the above simple estimation we conclude that the observable FLRW universe encompasses as may as  $\big( H_0^{-1}/H_{(m_e)}^{-1}\big)^3 \sim 10^{51}$ independent patches with the size $H_{(m_e)}^{-1}$. Demanding that as many as $10^{51}$ dS patches of the electron zero point energy to coherently cover the entire observable universe 
 is too strong a condition to be realistic.  Considering fields heavier than electron, the situation  becomes even worse, i.e, the ratio 
$ H_0^{-1}/H_{(m)}^{-1}$ becomes much larger.

\begin{figure}[t]
	\centering
	\includegraphics[ width=0.85\linewidth]{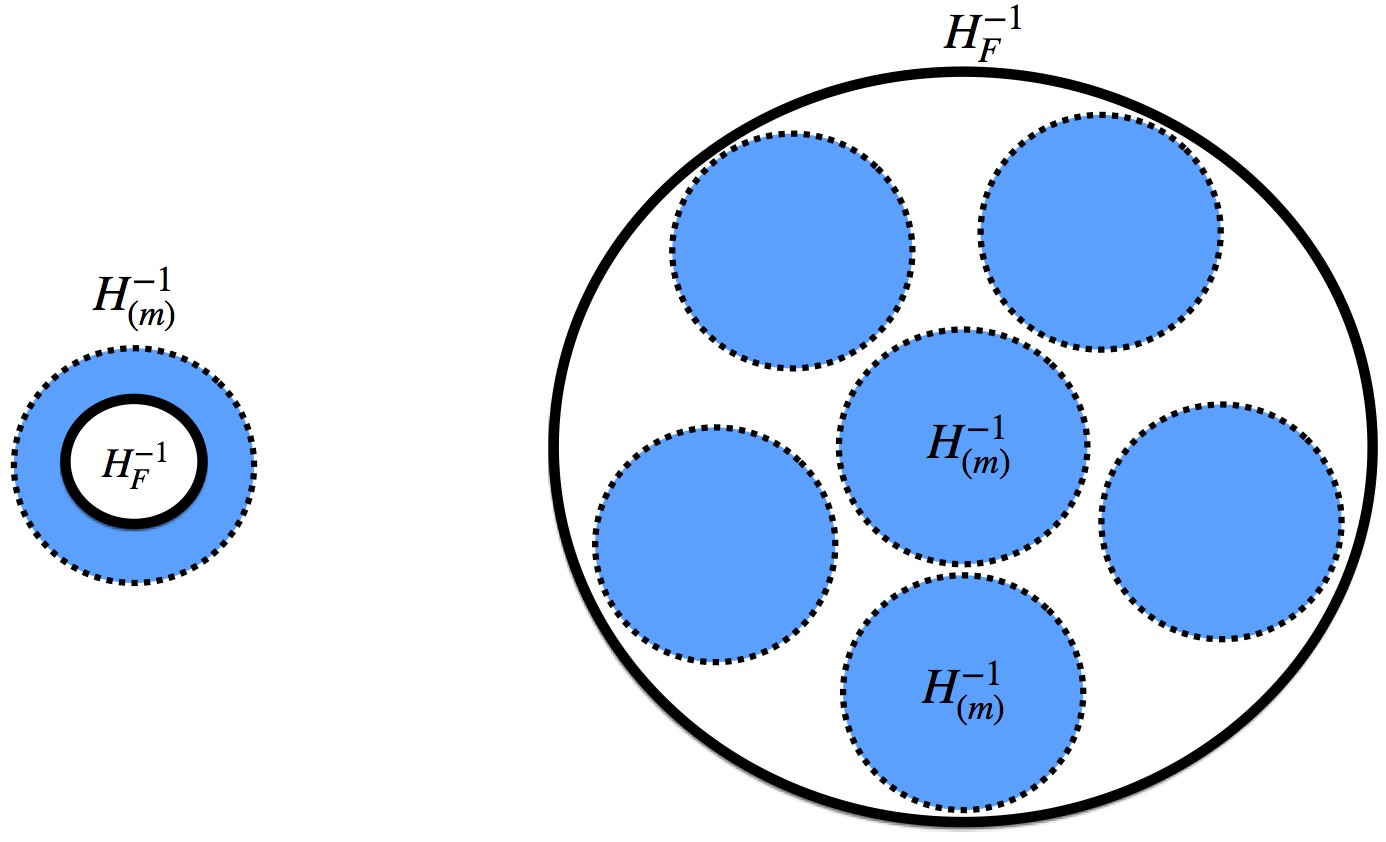}
	\vspace{1 cm}
	\caption{ A schematic view comparing the FLRW Hubble radius $H_F^{-1}$ with the dS horizon of  size $\Hm^{\, -1}$ associated with the zero point energy of the field with the mass $m$. Left: this is the case when the field is light so the  FLRW horizon is within a single dS patch of zero point energy. Right: this represents the case when the field is heavy so its dS horizon is much smaller than the FLRW 
	horizon such that a FLRW Hubble patch encompasses many small patches of zero point energy. }
\label{H0-Hm}
\end{figure}

The picture which emerges from the above discussions  is as follows. For heavy fields, the Hubble radius associated with a dS spacetime generated from the zero point energy is much smaller than the Hubble radius of the observable universe,  
$\Hm^{-1} \ll H_0^{-1}$. Therefore, one needs as many as 
\ba
\label{N-patch}
N_{\mathrm{patches}} \sim \Big( \frac{H_{(m)}}{H_0}\Big)^3 \sim \Big(\frac{m}{10^{-2} \mathrm{eV}} \Big)^6 \, 
\ea
tiny dS patches to cover the current observable universe. As these tiny patches are created quantum mechanically they  are uncorrelated so they can not provide a coherent energy density to be the origin of the observed cosmological constant (or dark energy).  This also provides an interesting resolution for the cosmological constant problem: considering the light fields. Specifically,  very light fields have much smaller  zero point energy but at the same time they have a very large Hubble radius which 
can encompass the entire observable universe in a single dS patch. From equation 
(\ref{N-patch}) we see that the (lightest) neutrino with the mass $m_\nu \sim 10^{-2} \mathrm{eV} $ has just the right scale to address the cosmological constant problem in which $H_0^{-1} \sim H_{(m_\nu)}^{-1}$.  
In this view  the entire observable universe currently is within a single dS patch  created by the zero point energy of the light neutrino field with the vacuum energy density $(10^{-3} \mathrm{eV})^4$, see the left panel of Fig. \ref{H0-Hm} for a schematic view.  Correspondingly,  the observed vacuum energy density today with the fractional energy density \cite{ Planck:2018vyg} $\Omega_\Lambda^{(0) } \sim 0.7$ is sourced by the zero point energy of the lightest neutrino.

 In the case of heavy fields, as the huge number of dS patches in Eq. (\ref{N-patch}) are uncorrelated, one expects that the full volume  between these regions to be highly inhomogeneous, for a schematic view see the right panel of Fig. \ref{H0-Hm}. 
 Consequently one expects that the variance in the energy density to be large, yielding to a large density contrast $\frac{\delta \rho_v(m)}{\rho_v(m)} \sim  1$. In the next Section we calculate the variance and the density contrast of 
the  vacuum zero point energy  which confirm the above conclusion. In addition, we calculate the correlation length associated with the zero point energy and show that the correlation length is  at the order of Compton radius $m^{-1}$.

However, before calculating the variance of the vacuum energy density, let us check the accumulated energy associated with the modes which are outside the correlation length, i.e. the modes which have crossed the quantum 
Compton radius with $k \leq m $.  
Denoting this accumulated energy by $\delta \rho_C(m)$ we obtain
 \ba
 \label{rhovC}
 \delta \rho_C(m) =  \frac{1}{2}\int_0^m \frac{d^3 \bfk}{( 2 \pi)^3} 
\sqrt{k^2 + m^2} = \frac{m^4}{32 \pi^2} \Big[  3 \sqrt{2} - \ln(1+ \sqrt{2}) \Big]
\sim \frac{6 m^4}{ 64 \pi^2} \, .
\ea
 Now we can compare the above value of the accumulated energy density 
 of the ``long modes"  with the supposedly background vacuum energy density $\rho_v(m)$ given in Eq. (\ref{cc-formula1}). 
 Assuming that there is no exponential hierarchy between $m$ and the renormalization scale $\mu$ 
 we obtain
 \ba
 \label{ratio-rhos}
 \frac{\delta \rho_C(m)}{\rho_v(m)} \sim 1 \, .
 \ea 
We conclude  that the accumulated energy density of the modes which are outside the correlation length $ m^{-1}$ is comparable to the background vacuum energy density.  This rings the bell that the background space can be highly inhomogeneous and the nearby dS patches can not provide a coherent background for the large scale universe.

\section{Density Contrast  and Correlation Length}
\label{variance-sec}

The quantum average of the vacuum zero point energy density  $\langle \rho_v \rangle=  \langle {\cal H} \rangle$ for the real scalar field is given in Eq. (\ref{cc-formula1}). Now for the moment suppose we neglect the contribution from the FLRW matter and radiation energy density and assume that $\rho_v$ is the dominant energy density in cosmic expansion.  In order for $\langle \rho_v \rangle$ to be viewed as a viable background energy density for the cosmic evolution we have to make sure that the variance in energy density is small, yielding to a small density contrast $\delta \rho_v/\langle \rho_v \rangle \ll 1$. Otherwise, a region with large density contrast $\delta \rho_v/ \langle \rho_v \rangle\sim1$ develops inhomogeneities and may even collapse to black holes.

In this section we calculate explicitly the variance $\delta \rho_v^2 \equiv \langle \rho_v^2 \rangle - \langle \rho_v \rangle^2$  and the density contrast $\delta \rho_v/\langle \rho_v \rangle $. First we present the analysis for the case of a real scale field and then go to the case of a Dirac fermion field. 


\subsection{Real Scalar Field }
\label{scalar-contrast}

First we consider the case of a real scalar field with the action (\ref{action-scalar}). 
Let us denote the three different contributions to energy density of the scalar field 
in Eq. (\ref{rho}) by $\rho_1, \rho_2$ and $\rho_3$ as follows,
\ba
\label{rho-i-def}
\rho_1\equiv  \frac{\dot \phi^2}{2} , \quad 
\rho_2 \equiv \frac{1}{2} \delta^{ij} \partial_i \phi \partial_j \phi , \quad
\rho_3 \equiv  \frac{m^2}{2} \phi^2 \, .
\ea 
To calculate $\langle \rho_i\rangle $ we go to Fourier space and expand the field in terms of the  mode functions as in Eq. (\ref{Fourier}).  Performing the momentum integrals using the dimensional regularization scheme  to regularize the UV divergences (for a sample analysis of dimensional regularization see Section \ref{zero-dS}) we obtain 
\ba
\langle \rho_1 \rangle =  \frac{1}{2} \int \frac{d^3 \bfq}{(2 \pi)^3}  
\frac{\omega (q)}{2} = \frac{m^4}{128 \pi^2} \ln \Big( \frac{m^2}{\mu^2}\Big)  \, ,
\ea
\ba
\label{rho2}
\langle \rho_2 \rangle =  \frac{1}{2} \int \frac{d^3 \bfq}{(2 \pi)^3}  
\frac{\bfq^2}{2 \omega (q)} = \frac{-3 m^4}{128 \pi^2} \ln \Big( \frac{m^2}{\mu^2}\Big) \, ,
\ea
\ba
\label{rho3}
\langle \rho_3 \rangle =  \frac{1}{2} \int \frac{d^3 \bfq}{(2 \pi)^3}  
\frac{m^2}{2 \omega (q)}=  \frac{4 m^4}{128 \pi^2} \ln \Big( \frac{m^2}{\mu^2}\Big) \, .
\ea
Curiously, note that while all three operators $\rho_1, \rho_2$ and $\rho_3$ 
$\rho_2$ are positive definite and Hermitian operators but the expectation value 
of $\rho_2$ has opposite sign compared to  those of  $\rho_1$ and $\rho_2$.  
Furthermore, noting that $\langle \rho \rangle = \langle \rho_1 \rangle + 
\langle \rho_2 \rangle + \langle \rho_3 \rangle $, 
the following relations hold which will be useful later on
\ba
\label{rho123}
\langle \rho \rangle = 2 \langle \rho_1 \rangle=  -\frac{2}{3}  \langle \rho_2 \rangle
= \frac{1}{2}  \langle \rho_3 \rangle  \, .
 \ea

Now to calculate $\delta \rho^2$ we have 
\ba
\delta \rho^2  &=&  \langle \rho_1^2 \rangle +  \langle \rho_2^2 \rangle+ \langle \rho_3^2 \rangle + \langle \rho_1 \rho_2 \rangle  +  \langle \rho_2 \rho_1 \rangle +
\langle \rho_1 \rho_3 \rangle+ \langle \rho_3 \rho_1 \rangle+
\langle \rho_2 \rho_3 \rangle+\langle \rho_3 \rho_2 \rangle  \nonumber\\
&-&  \Big( \langle \rho_1 \rangle^2 +  \langle \rho_2 \rangle^2+ \langle \rho_3 \rangle^2 \Big)  
- 2   \langle \rho_1 \rangle  \langle \rho_2 \rangle -2  \langle \rho_1 \rangle  \langle \rho_3 \rangle -2  \langle \rho_2  \rangle  \langle \rho_3 \rangle  \, .
\ea
Note that since $\dot \phi$ and $\phi$ do not commute we have considered all possible orderings of the operators. However, as we shall see, these orderings  do not matter.

First we note that $\phi$ is a free Gaussian field so using the standard Wick theorem (or by direct calculations) one can show that
\ba
\label{rho-squared1}
\langle \rho_i^2 \rangle = 3 \langle \rho_i \rangle^2 \, , \quad \quad  i=1, 3. 
\ea 
However, the calculation of $\langle \rho_2^2 \rangle$ is somewhat tricky. Performing various contractions we obtain
\ba
\label{rho-squared2}
\langle \rho_2^2 \rangle &=&  \langle \rho_2 \rangle^2 + (2)\frac{1}{4} \int \frac{d^3 \bfq_1}{(2 \pi)^3 2 \omega(q_1)} \frac{d^3 \bfq_2}{(2 \pi)^3 2 \omega(q_2)}  
(\bfq_1 \cdot \bfq_2)^2 \nonumber\\
&=&  \langle \rho_2 \rangle^2 + (2\times \frac{1}{3})   \langle \rho_2 \rangle^2 \, ,
\ea
in which the factor $1/3$ originated from an integral of the form $\int_{-1}^{+1} d(\cos \theta) \cos^2 \theta$.

For the mix correlations we calculate one case for illustration and the rest would be similar.  For this purpose let us calculate $\langle \rho_1 \rho_2 \rangle$. To simplify the analysis note that due to translation invariance we can simply set $\bfx=0$, obtaining
\ba
\langle \rho_1 \rho_2 \rangle  &=&  \frac{i^4}{4} \int  \frac{d^3 \bfq_i}{(2\pi)^{\frac{3}{2}}} \frac{\omega(q_1) \omega(q_2) \bfq_3\cdot \bfq_4}{2 \omega(q_i)} \langle 0| (- a_{\bf q_1} + a_{\bfq_1}^\dagger)
(- a_{\bf q_2} + a_{\bfq_2}^\dagger)  ( a_{\bf q_3} - a_{\bfq_3}^\dagger)
( a_{\bf q_4} - a_{\bfq_4}^\dagger)   | 0 \rangle  \nonumber \\
  &=&\frac{1}{4} \int   \frac{d^3 \bfq_1} {(2\pi)^{3} 2 \omega(q_1)}  \frac{d^3 \bfq_2} {(2\pi)^{3}  2 \omega(q_2)}
\Big[  2 \omega(q_1) \omega(q_2) \bfq_1 \cdot \bfq_2 + \omega(q_1)^2 \bfq_2^2
\Big] \, .
\ea
On the other hand, due to rotation invariance the integral over $\bfq_1 \cdot \bfq_2$
vanishes and we obtain 
\ba
\langle \rho_1 \rho_2 \rangle = \frac{1}{4} \int \frac{d^3 \bfq_1}{(2 \pi)^3} \frac{d^3 \bfq_2}{(2 \pi)^3} \frac{\omega (q_1)}{2} \frac{\bfq_2^2}{2 \omega(q_2)} 
= \langle \rho_1 \rangle   \langle \rho_2 \rangle \, .
\ea
Similarly, we obtain
\ba
\langle \rho_2 \rho_3 \rangle = \langle \rho_2 \rangle   \langle \rho_3 \rangle \, .
\ea
Since $\dot \phi$ and $\phi$ do not commute, the calculation for $\langle \rho_1 \rho_3 \rangle$ is a bit different, yielding
\ba
\langle \rho_1 \rho_3 \rangle = \langle \rho_1 \rangle   \langle \rho_3 \rangle 
- \frac{m^2}{2} \Big( \int   \frac{d^3 \bfq_1} {(2\pi)^{3}}  \Big)^2 \, .
\ea
Regularizing  the last term above which is UV divergent by absorbing into counter terms,  we obtain
\ba
\langle \rho_1 \rho_3 \rangle =  \langle \rho_1 \rangle   \langle \rho_3 \rangle 
= \langle \rho_3 \rho_1 \rangle \, .
\ea
Combining all, we obtain
\ba
\label{delta-rho}
\delta \rho^2  &=& 2 \Big( \langle \rho_1 \rangle^2 +  \langle \rho_2 \rangle^2+ \langle \rho_3 \rangle^2 \Big)  - \frac{4}{3} \langle \rho_2 \rangle^2
 \nonumber\\
&=&  10   \langle \rho \rangle^2
\, ,
\ea
where to obtain the final result the relations (\ref{rho123}) have been used. 

From the above expression, the density contrast (taking both signs after the square root) is obtained to be   
\ba
\label{var-sqrt10}
\frac{\delta \rho_v}{ \langle \rho_v \rangle}  = \pm \sqrt{10} \, .
\ea
This is an interesting result. As argued before, the energy density of the background 
constructed from the zero point energy is very inhomogeneous.  This is in agreement with the intuition that a large number of incoherent patches of zero point energy can not cover the entire observable universe. Also note that both signs are allowed so we can have either over-density or under-density. Specifically,  the quantity $\langle \rho \rangle + \delta \rho $ can take either signs  so some regions of spacetime can be in dS-type and some other parts  AdS-type.  The parts of the spacetime which are in AdS-type may collapse due to instabilities.  

The above conclusion that $\langle \rho \rangle + \delta \rho$ can be negative is based on the hidden assumption that the statistical distributions of $\rho$ is mostly symmetric around its average value $\langle \rho \rangle$ so  a large variance can 
yield to a negative local value of $\langle \rho \rangle + \delta \rho$. This is certainly the case for a Gaussian distribution while for other distribution like Poisson distribution it may not be the case. Therefore,  the nature of the distribution of $\rho$ is an important question. In our setup  with a large uncorrelated dS patches of heavy fields  created quantum mechanically  from the vacuum it may be reasonable 
to assume that $\rho$ follows a  Normal (i. e. Gaussian) distribution. But it  would be interesting to calculate the distribution of $\rho$ from the first principle. In the following analysis  we assume a 
Gaussian-type distribution such that  $\langle \rho \rangle + \delta \rho$ can become negative for a large variance.

In the above analysis we have neglected the conventional matter and radiation 
energy density of the FLRW universe, $\rho_F$.  Now, let us look at the total energy density from the combination of the FLRW matter-radiation $\rho_F$ and the  vacuum energy density, defined via $\rho_T = \rho_F + \rho_v$ in which $\rho_v$ is the vacuum energy density given in Eq. (\ref{cc-formula1}).
Note that $\rho_F$ is independent of the zero point energy so its quantum expectation is trivial, i.e. $\langle \rho_F \rangle = \rho_F$. Correspondingly,  
$\langle \rho_T \rangle = \rho_F + \langle \rho_v \rangle$ 
and $\delta \rho = \delta \rho_v$ in which $\delta \rho_v$ is as given in 
Eq. (\ref{delta-rho}).  Therefore, the   total density contrast 
 is given by
\ba
\label{fraction} 
\frac{\delta \rho_T}{ \langle \rho_T \rangle} = \frac{\delta \rho_v}{\rho_F 
+ \langle \rho_v \rangle} =  \pm \sqrt{10}  \frac{  \langle \rho_v \rangle}{\rho_F 
+ \langle \rho_v \rangle} \, .
\ea
Demanding that the amplitude of the total density contrast to be small, 
$| \frac{\delta \rho_T}{ \langle \rho_T \rangle} | <1 $,  we obtain
\ba
\label{total-var}
 \langle \rho_v \rangle < \Big( \sqrt{10} -1\Big)^{-1}   \rho_F 
 \sim \frac{ \rho_F}{2} \, .
\ea
This means that in order for the background consisting of matter-radiation plus zero point energy to be stable for cosmological expansion one requires that the vacuum energy should  be less than the FLRW matter-radiation energy density. 

Motivated by the above analysis, it would be instructive to calculate the fractional contrast in  ${\cal K} \equiv \rho+ 3 p$ as the quantity ${\cal K}$ (the strong energy condition with ${\cal K}\ge 0$)  plays key roles in gravitational collapse and black hole formations
in Hawking-Penrose theorem.   One can check that the vacuum  pressure is given by
\ba
p_v = \rho_1 - \frac{\rho_2}{3} - \rho_3 \, ,
\ea
so we have 
\ba
\label{T-form}
 {\cal K}= 4 \rho_1 - 2 \rho_3 \, .
\ea
Now, repeating the same steps as in the case of $\rho$, one can
show that  $\big \langle {\cal K}  \big \rangle = - 2 \langle \rho_v \rangle$
and $\big\langle {\cal K}^2  \big\rangle = 44 \langle \rho_v \rangle^2$. Combining these two results we obtain
\ba
\delta {\cal K}^2 \equiv  \big \langle{\cal K}^2  \big \rangle - \langle {\cal K} \rangle^2= 
40 \langle \rho_v \rangle^2 \, ,
\ea
and
\ba
\frac{\delta {\cal K}}{\langle {\cal K} \rangle } =  \pm \sqrt{10 } \, .
\ea
This suggests  that the local curvature of the spacetime fluctuates rapidly in which  some regions are expected to become AdS type and may collapse to form black holes.

Having calculated the variance and the density contrast, it is also instructive to calculate the correlation length of the energy density. The correlation length is related to the connected part of the correlation function as follows. The two point correlation function is given 
 by $\langle \rho(\bfx) \rho(\bf{y})\rangle$. Due to translation invariance the correlation function depends only on $\bfx- \bf{y}$ so we simply set $\bf{y}=0$. The connected part of the two point correlation function is given by
 \ba
{ \big \langle \rho(\bfx) \rho(\bf{0}) \big \rangle}_c \equiv  \big\langle \rho(\bfx) \rho({\bf{0}}) \big\rangle
 - \langle {\rho^2}\rangle \, .
 \ea 
 Performing the analysis similar to the case of variance one can calculated ${ \langle \rho(\bfx) \rho(\bf{0})\rangle}_c$. The analysis is somewhat involved. However, what we are interested in is the correlation length $\xi$ which controls the  exponential fall off of the connected correlation function on large distance, 
 \ba
 { \big\langle \rho(\bfx) \rho(\bf{0}) \big\rangle}_c \rightarrow e^{-\frac{r}{\xi} }\, ,   \quad  \quad 
 ( r \rightarrow \infty) \, .
 \ea
To estimate $\xi$ let us look at ${\langle \rho_3(\bfx) \rho_3(\bf{0})\rangle}_c$ which is easier. Upon performing the various contractions we obtain 
\ba
{\big\langle \rho_3(\bfx) \rho_3(\bf{0}) \big\rangle}_c = \frac{m^{4}}{2} \Big(  \int \frac{d^3 \bfq}{(2 \pi)^3 2 \omega(q)} e^{-i \bfq \cdot \bfx} \Big)^2 \, .
\ea
The above integral is well-known in QFT which yields \cite{Weinberg:1995mt}
\ba
{\big\langle \rho_3(\bfx) \rho_3(\bf{0}) \big\rangle}_c = \frac{m^8}{32 \pi^4} 
\Big( \frac{K_1(m r)}{mr}\Big)^2 \, ,
\ea
in which $K_1(x)$ is the modified Bessel function. Using the asymptotic behaviour 
$K_1(x) \sim e^{-x}$ we obtain 
$ {\langle \rho_3(\bfx) \rho_3(\bf{0})\rangle}_c \sim  e^{-2 mr} $ yielding the correlation length $\xi = 1/(2m)$. This is somewhat expected as $m$ is the only mass scale relevant for this free QFT. Calculating the two point correlation functions for  other components of the energy density yield  the same exponential behaviour for large $r$.

Now having the correlation length at hand, we can compare the Hubble radius of zero point energy $\Hm^{-1}$ and $\xi$, obtaining
\ba
\frac{\xi}{\Hm^{-1}} \sim \frac{\Hm}{m} \sim \frac{m}{M_P} \ll 1. 
\ea
This is inline with our intuitive argument in previous section. The correlation length of the zero point energy density is much smaller than the Hubble radius of each dS patch so the dS patches are practically uncorrelated. As such, one can not cover the entire FLRW horizon by a vast number of uncorrelated dS patches (like the right panel of Fig. \ref{H0-Hm}) which are created quantum mechanically 
and yet expect them to behave as  a  uniform cosmological constant.

As the above discussions suggest, the dS spacetime created purely from the vacuum zero point energy with $\delta \rho_v/  \langle  \rho_v \rangle \sim \pm 1$
is unstable under perturbations. Does this suggest that the locally formed AdS regions collapse to black hole?  We can not be sure about the answer to this question (see Section \ref{implications-sec} for further discussions/speculations). But suppose that some regions may collapse to form black holes. It is therefore helpful to compare the Jeans length of the corresponding gravitational instability and the horizon size of the supposedly formed black holes. Assuming  a dS patch of size $\Hm^{-1}$ to collapse into a black hole, the mass of the corresponding black hole is at the order 
$M \sim \rho_v \Hm^{-3} \sim M_P^2 \Hm^{-1} \sim M_P^3/m^2$. This yields to a 
 Schwarzschild radius $r_S \sim \Hm^{-1}$. In other words, the Schwarzschild radius is nothing but the horizon radius of the dS patch. On the other hand, the Jeans length of perturbations is given by $\lambda_J \sim c_s/\sqrt{G \rho_v}$ in which $c_s$ is the sound speed of perturbations. In our picture with 
 $\delta \rho_v/ \langle \rho_v \rangle \sim \pm 1$, parts of spacetime fragment to AdS-type while other pars are in dS-type (though with a higher Hubble expansion rate than the background dS).  Assuming that the sound speed of perturbations are not exponentially small, the 
 Jeans length is at the order $\lambda_J \sim \sqrt{M_P^2/\rho_v} \sim \Hm^{-1} \sim r_S $. Interestingly, we conclude that the Jeans length of gravitational instability is at the same order as the black hole horizon. This may support the conclusion that  the dS spacetime constructed from the vacuum zero point energy is unstable to fragmentations and formation of black holes of size $\Hm^{-1}$. However, as just mentioned above, the situation is more complicated and we leave it as an open question if locally formed AdS regions from large quantum 
 fluctuations can collapse to form black hole. 
 
Before closing this subsection, there is an important comment in order. In our analysis we have not taken into account the backrecations of strong inhomogeneities on the background geometry. As we have seen, for a pure vacuum dominated energy density, we have $\delta \rho_v/\rho_v \sim 1$. Correspondingly, this induces strong inhomogeneities in geometry. At the start, assuming a uniformly distributed vacuum energy, we have assumed a homogenous and isotropic FLRW 
background determined by the induced Hubble expansion rate $\Hm$. In the presence of strong inhomogeneities, this assumption is violated and one has to consider a general inhomogeneous and anisotropic background to properly take into account the strong inhomogeneities from the vacuum fluctuations. This is a difficult and open question.  On the other hand, as we have stressed above, to have a consistent background, we require a classical source of energy density, $\rho_F$,  such that the total density contrast $\delta \rho_v/\rho_T $ remains under perturbative control. However, this assumption breaks down  when $\rho_F$ falls off as the universe expands and when 
$\rho_F \sim \rho_v$. At this stage the inhomogeneities from vacuum fluctuations can not be neglected. This will bring important questions which we will come back in Section \ref{implications-sec}.


\subsection{Dirac Fermion Field } 
 Here we present the analysis of variance and  density contrast 
 for a Dirac fermion field $\Psi$. In what follows we follow the notations of Weinberg \cite{Weinberg:1995mt}. 
 
 Expressing the mode function in Fourier space, we have 
 \ba
 \label{Psi-eq}
 \Psi = \int \frac{d^3 \bfq}{(2 \pi)^{\frac{3}{2}}} \sum_\sigma\Big[  u (\bfq, \sigma) e^{i q\cdot x} a(\bfq, \sigma) +  v (\bfq, \sigma) e^{-i q\cdot x} b^\dagger(\bfq, \sigma)
 \Big] \, ,
 \ea
in which $a(\bfq, \sigma)$ and $b(\bfq, \sigma)$ are the annihilation operators for the ``particle" and  ``antiparticle", $\sigma=\pm\frac{1}{2}$ are two spin 
polarizations while 
$ u (\bfq, \sigma)$ and $ v (\bfq, \sigma)$ are the corresponding mode functions. As usual, $a(\bfq, \sigma)$ and $b(\bfq, \sigma)$ are anti commuting operators in which  $\{ a(\bfq, \sigma) , a^\dagger(\bfq', \sigma')  \}= \delta_{\sigma \sigma'}\delta (\bfq- \bfq')$ and $\{ b(\bfq, \sigma) , b^\dagger(\bfq', \sigma')  \}= \delta_{\sigma \sigma'}\delta (\bfq- \bfq')$ while the rest of anti commutators are zero. 

The action of the field with the mass $m$ is given by
\ba
\label{action-fermion}
S= - \int d^4 x \,   \bar{\Psi} (\gamma^\mu \partial_\mu + m) \Psi  \, ,
\ea
in which $\gamma^\mu$ are the Dirac matrices satisfying the anti commutation relation
\ba
\{ \gamma^\mu, \gamma^\nu \} = 2 \eta^{\mu \nu} \, ,
\ea
while $\bar \Psi\equiv \Psi^\dagger \beta $ in which the matrix $\beta$ is given by $\beta \equiv i \gamma^0$. 

The mode functions $ u (\bfq, \sigma)$ and $ v (\bfq, \sigma)$ satisfy the following relations 
\ba
\label{rel1}
\sum_\sigma u_\mu (\bfq, \sigma) \bar u_\nu(\bfq, \sigma) = \Big( \frac{-i \gamma^\mu q_\mu + m}{2 q^0}
\Big)_{\mu \nu} \, ,
\ea
and
\ba
\label{rel2}
\sum_\sigma v_\mu (\bfq, \sigma) \bar v_\nu(\bfq, \sigma) = \Big( \frac{-i \gamma^\mu q_\mu - m}{2 q^0}
\Big)_{\mu \nu} \, ,
\ea
in which $q^0 = \sqrt{\bfq^2 + m^2}$ is the energy of the particle.  Furthermore, the following relation is useful as well:
\ba
\label{rel3}
\sum_\sigma \bar v(\bfq, \sigma) \gamma^0 v(\bfq , \sigma) = -2 i \, .
\ea

Finally, the energy density $\rho= T_{00}$ is given by
\ba
\label{rho-fermion}
\rho= -\frac{1}{2} \bar \Psi \gamma^0 \partial_0 \Psi +  \frac{1}{2} \partial_0 \bar \Psi \gamma^0  \Psi  \, .
\ea
It is convenient to define the above two terms via $\rho_1 \equiv -\frac{1}{2} \bar \Psi \gamma^0 \partial_0 \Psi$ and  $\rho_2 \equiv \frac{1}{2} \partial_0 \bar \Psi \gamma^0  \Psi$ (these values of $\rho_i$ should not be confused with $\rho_i$ in the case of a real scalar field studied in previous subsection).  
One can check that $\rho_2 = \rho_1^\dagger $ so 
$\rho = \rho_1 + \rho_1^\dagger$. This shows, as expected, that $\rho$ is a Hermitian operator. 

As a warm up, it is helpful to calculate $\langle \rho \rangle$ from Eq. (\ref{rho-fermion}). We obtain $\langle \rho \rangle = \langle \rho_1 \rangle + \langle \rho_1 \rangle^*$, so we only need to calculate  $\langle \rho_1 \rangle$. Using the decomposition of the field in terms of the mode functions in Eq. (\ref{Psi-eq}), we have
\ba
\label{rho1-av}
\langle \rho_1 \rangle &=& -\frac{1}{2} \langle 0 \big| \bar \Psi \gamma^0 \partial_0 \Psi  \big | 0 \rangle   \nonumber\\
&=& -\frac{i}{2} \int \frac{d^3 \bfq}{(2 \pi)^3} \sum_\sigma \bar v(\bfq, \sigma) \gamma^0 v(\bfq , \sigma) q^0 \nonumber\\
&=& - \int \frac{d^3 \bfq}{(2 \pi)^3} q^0 \, .
\ea
As a result,
\ba
\label{rho-av}
\langle \rho \rangle = 2 \langle \rho_1 \rangle = -4  \int \frac{d^3 \bfq}{(2 \pi)^3} \frac{q^0}{2} \,  = -4 \Big[  \frac{m^4}{64 \pi^2} \ln \Big( \frac{m^2}{\mu^2}\Big) \Big] \, .
\ea
The above result confirms the factor 4  since we have four independent degrees of freedom in the Dirac field. In addition, compared to the case of a real scalar field given in Eq. (\ref{rhov}) we have an overall minus sign which is the hallmark of the fermionic field. 

To calculate $\langle \rho^2 \rangle$ note that $\langle \rho_2 \rangle= \langle \rho_1 \rangle^*$,  $\langle \rho_2^2 \rangle= \langle \rho_1^2 \rangle^*$ and
$\langle \rho_1 \rho_2 \rangle= \langle \rho_1 \rho_2 \rangle^* =  \langle \rho_2 \rho_1 \rangle $ which will be handy in the following analysis.    Furthermore, in the analysis below we show that $ \langle \rho_1^2 \rangle$ and $\langle \rho_1 \rho_2 \rangle$ are real, so using the above mentioned relations, we obtain 
\ba
\label{rho2-fermion0}
\langle \rho^2 \rangle = 2 \langle \rho_1^2 \rangle  
+ 2 \langle \rho_1 \rho_2 \rangle  \, .
\ea
We calculate each term in turn. Due to translation invariance we set $x^\mu=0$
without loss of generality. 

Starting with $\langle \rho_1^2 \rangle$ we have 
\ba
\langle \rho_1^2 \rangle = \frac{1}{4}  \langle 0 \big|  \big( \bar \Psi \gamma^0 \partial_0 \Psi \big) \big(  \bar \Psi \gamma^0 \partial_0 \Psi \big)
\big | 0 \rangle \, .
\ea
Using the decomposition in terms of mode functions from Eq. (\ref{Psi-eq}) and performing all the possible contractions involving the creation and annihilation operators, we end up with two types of contributions in  $\langle \rho_1^2 \rangle$. Denoting these contributions as term  $A$ and term  $B$, we have $\langle \rho_1^2 \rangle \equiv A + B$ in which
\ba
\label{A-term}
A\equiv \frac{1}{4} \sum_{\sigma_1 \sigma_2}  \int \frac{d^3 \bfq_1}{(2 \pi)^3}
\frac{d^3 \bfq_2}{(2 \pi)^3} \,   q_1^0  \, q_2^0 \, \,  \bar v(\bfq_1 , \sigma_1) \gamma^0 u(\bfq_2 , \sigma_2)  \, \,  \bar u(\bfq_2 , \sigma_2) \gamma^0 v(\bfq_1 , \sigma_1) \, ,
\ea 
and
\ba
\label{B-term} 
B\equiv -\frac{1}{4} \sum_{\sigma_1 \sigma_2}  \int \frac{d^3 \bfq_1}{(2 \pi)^3}
\frac{d^3 \bfq_2}{(2 \pi)^3} \,   q_1^0  \, q_2^0 \, \,  \bar v(\bfq_1 , \sigma_1) \gamma^0 v(\bfq_1 , \sigma_1)  \, \,  \bar v(\bfq_2 , \sigma_2) \gamma^0 v(\bfq_2 , \sigma_2) \, .
\ea 

Using the relations (\ref{rel1}) and (\ref{rel2}) we obtain
\ba
\label{Aterm-2}
A   &=&   \frac{1}{4}  \int \frac{d^3 \bfq_1}{(2 \pi)^3}
\frac{d^3 \bfq_2}{(2 \pi)^3} \, \big( m^2 - q_1^0 q_2^0 \big)  \nonumber\\
&=& \frac{m^2}{4} \Big (\int \frac{d^3 \bfq}{(2 \pi)^3} \Big)^2 - \frac{1}{4} 
\langle \rho_1 \rangle^2 \, ,
\ea
where to obtain the last result, Eq. (\ref{rho-av})  for $\langle \rho_1 \rangle$ has been used. 

Similarly, using the relation (\ref{rel3}), for the $B$ term we obtain
\ba
\label{Bterm-2}
B= \Big (\int \frac{d^3 \bfq}{(2 \pi)^3} q^0 \Big)^2 = \langle \rho_1 \rangle^2 \, .
\ea

Combining the above two contributions, we obtain
\ba
\label{rho1-av}
\langle \rho_1^2 \rangle =  A + B =  \frac{m^2}{4} \Big (\int \frac{d^3 \bfq}{(2 \pi)^3} \Big)^2 + \frac{3}{4}  \langle \rho_1 \rangle^2 \, .
\ea
As promised, $\langle \rho_1^2 \rangle$ is real so we do not need to calculate
$\langle \rho_2^2 \rangle$ since $\langle \rho_2^2 \rangle = \langle \rho_1^2 \rangle^* = \langle \rho_1^2 \rangle$. Furthermore, as in the case of scalar field, the first term above is UV divergent and has to be absorbed into counter terms via regularization so after regularization we obtain $\langle \rho_1^2 \rangle \cong  \frac{3}{4}  \langle \rho_1 \rangle^2$. 

Now we calculate $\langle \rho_1 \rho_2 \rangle$ which is
\ba
\label{rho12-av}
\langle \rho_1 \rho_2 \rangle = -\frac{1}{4}  \langle 0 \big|  \big( \bar \Psi \gamma^0 \partial_0 \Psi \big) \big(  \partial_0 \bar \Psi \gamma^0  \Psi \big)
\big | 0 \rangle \, .
\ea 
Like in the previous case, after performing all contractions involving the creation and the annihilation operators we have two different contributions in $\langle \rho_1 \rho_2 \rangle$. Incidentally, one of this contribution is the $B$ term above while the 
other contribution is new, denoted by the $C$ term so $\langle \rho_1 \rho_2 \rangle = B + C$ in which  
\ba
\label{C-term}
C \equiv   -\frac{1}{4} \sum_{\sigma_1 \sigma_2}  \int \frac{d^3 \bfq_1}{(2 \pi)^3}
\frac{d^3 \bfq_2}{(2 \pi)^3} \,   (q_2^0)^2 \, \,  \bar v(\bfq_1 , \sigma_1) \gamma^0 u(\bfq_2 , \sigma_2)  \, \,  \bar u(\bfq_2 , \sigma_2) \gamma^0 v(\bfq_1 , \sigma_1) \, .
\ea
Note that while the $C$ term looks very similar to the $A$ term, but they are different as the latter has the product $q_1^0 q_2^0$ while the former has 
$(q_2^0)^2$ inside the double integrals. 

Using the relations (\ref{rel1}) and (\ref{rel2}) we obtain
\ba
\label{C-term2}
C=  -\frac{m^2}{4} \Big (\int \frac{d^3 \bfq_1}{(2 \pi)^3} \frac{1}{q_1^0} \Big) 
\Big (\int \frac{d^3 \bfq_2}{(2 \pi)^3} {q_2^0} \Big)
+ \frac{1}{4} \int \frac{d^3 \bfq_1}{(2 \pi)^3}  \int \frac{d^3 \bfq_2}{(2 \pi)^3} {(q_2^0)^2}  \, .
\ea

Combining the two contributions $B$ and $C$, we obtain
\ba
\langle \rho_1 \rho_2 \rangle =  \langle \rho_1 \rangle^2 +\frac{m^2}{4} \Big (\int \frac{d^3 \bfq}{(2 \pi)^3} \frac{1}{q^0} \Big)   \langle \rho_1 \rangle +
\frac{1}{4} \int \frac{d^3 \bfq_1}{(2 \pi)^3}  
\int \frac{d^3 \bfq_2}{(2 \pi)^3} {(q_2^0)^2}  \, .
\ea
The last term above is again UV divergent and has to be absorbed into counter terms so we keep the first two terms above in the regularized 
$\langle \rho_1 \rho_2 \rangle$.

Having calculated $\langle \rho_1^2 \rangle$ and $\langle \rho_1 \rho_2 \rangle$ 
and discarding the two UV divergent terms as discussed above, 
 from Eq. (\ref{rho2-fermion0})  $\langle \rho^2 \rangle$ is calculated to be 
 \ba
 \label{rho-final0}
 \langle \rho^2 \rangle  = \frac{7}{2} \langle \rho_1 \rangle^2 + \frac{m^2}{2} \Big (\int \frac{d^3 \bfq}{(2 \pi)^3} \frac{1}{q^0} \Big)   \langle \rho_1 \rangle \, .
 \ea
On the other hand, the integral in the second term above is already  calculated in
the case of the real scalar field in Eq. (\ref{rho3}) which, combined with  Eq.
(\ref{rho-av}), can be expressed in terms of $\langle \rho_1 \rangle$ as
\ba
\frac{m^2}{2} \Big (\int \frac{d^3 \bfq}{(2 \pi)^3} \frac{1}{q^0} \Big)   = -2  \langle \rho_1 \rangle \, .
\ea

Now combining all the terms, we obtain
\ba
 \label{rho-final}
 \langle \rho^2 \rangle  = \frac{3}{2} \langle \rho_1 \rangle^2  =  \frac{3}{8} \langle \rho \rangle^2 \, .
 \ea
Finally, the variance  $ \delta \rho^2 =  \langle \rho^2 \rangle - \langle \rho \rangle^2$
is obtained to be
\ba
\label{var-fermion} 
\delta \rho^2 = -\frac{10}{16}  \langle \rho \rangle^2 \, .
\ea
Surprisingly, the variance is negative! This may look absurd. However, we already encountered seemingly negative value for the expectation values of positive definite operators, such as the expectation value of  $\langle \rho_2 \rangle$ in Eq. (\ref{rho2}) for the case of real scalar field.  This phenomena may be attributed to regularization which absorbs positive infinity terms leaving the finite possible negative terms. 

The amplitude of the zero point density perturbation is obtained to be 
\ba
\label{density-contrast}
\big{|} \frac{\delta \rho_v}{\rho_v} \big{|} = \frac{\sqrt{10}}{4} \, ,
\ea
in which by $\big{|} \frac{\delta \rho_v}{\rho_v} \big{|}$ we mean the amplitude 
with the minus sign absorbed. Curiously, the factor $\sqrt{10}$ is common as in the case of a real scalar field, see Eq. (\ref{var-sqrt10}).   The factor $\frac{1}{4}$ may be attributed to the four degrees  of freedom encoded in Dirac fermion field. Furthermore, calculating the variance of ${\cal K}$ (defined like in the case of scalar field) we obtain 
$|\frac{\delta {\cal K}}{\langle {\cal K} \rangle } | = \frac{\sqrt{10}} {4}$. 
 
 As in the case of scalar field we see that the amplitude of density contrast associated  to the  zero point energy of the fermionic field is at the oder unity. As before, we conclude that the vacuum zero point energy by itself can not be the dominant source of the cosmic energy density. One requires a dominant (classical) source of energy density like the matter-radiation energy density $\rho_F$ with $\rho_F > \rho_v$ at each stage in cosmic expansion history to obtain a stable cosmological background.  
 
 Finally, calculating the correlation length $\xi$ of the zero point energy, one can check that, as in the case of real scalar field, $\xi \sim m^{-1}$ as expected.

\section{Zero Point Energy in dS Background}
\label{zero-dS}

The vacuum zero point  energy density Eq. (\ref{cc-formula1})
is calculated assuming a flat  background. As discussed before, one expects this result to hold true for a curved background as well. This is because general relativity is a local theory and the combination of the Lorentz invariance and the equivalence principle guarantees the validity of Eq. (\ref{cc-formula1}) even in a curved field space. Having said this, it is a non-trivial exercise to see how this works 
in practice. Specifically, the accumulated zero point energy yielding to Eq. (\ref{cc-formula1}) generate a dS background. At the same time, one has to solve the mode function $ \phi(x^\mu)$ in this dS spacetime and then construct 
$\langle T_{00} \rangle$. Technically speaking, it seems non-trivial how Eq. (\ref{cc-formula1}) emerges as the final result. This is more intriguing for a dS spacetime noting that a dS spacetime is characterized by the size of its horizon, $\Hm^{-1}$ in our notation, while  Eq. (\ref{cc-formula1}) is indifferent to this length scale.   

In this Section we solve the mode function for a real scalar field 
in a dS background and calculate the vacuum zero point energy. Note the non-trivial 
step that we have to calculate the mode function in a dS space which itself 
is created from the vacuum zero point energy.  In other words, {\it there is no}  background dS space other than what would be generated from the vacuum zero point energy in the first place. This resembles a strong backreaction problem where we have to create a background out of quantum perturbations and at the same time solve the mode function in the geometry of the created background. Then we have to regularize the energy density to make sure the resultant value of $\langle \rho \rangle$ agrees with its original value in the supposedly created background.  
This seems like moving in a loop!  For relevant works  see also Refs. \cite{Martin:2012bt, Onemli:2002hr, Moreno-Pulido:2020anb, Moreno-Pulido:2021jmn} and \cite{ Birrell:1982ix} (Chapter 6) who studied the  analysis of zero point energy in a {\it given} curved background. For earlier works concerning the renormalized stress energy tensor  see \cite{Dowker:1975tf, Bunch:1978yq, Bunch:1978yw, Davies:1977ze }

To treat the regularization properly, we employ dimensional regularization for a free scalar field in a curved field space with the spacetime dimension $d$, with the action
\ba
S =\int d^{d} x \sqrt{-g} \Big[ -\frac{1}{2} \partial_\mu \phi \partial^\mu \phi - \frac{m^2}{2} \phi^2 \Big].
\ea
The background metric is given by a FLRW type metric in $d$ dimension,
\ba
ds^2 = - dt^2 + a(t)^2  d \bfx^2 \, ,
\ea
in which $a(t)$ is the scale factor. 

Using the conformal time $d\tau= dt/a(t)$ and defining the canonically normalized field $\sigma$ via $\sigma \equiv a^{\frac{d-2}{2}} \, \phi$, the action takes the following form
\ba
\label{avtion-canonocal}
S= \frac{1}{2} \int d^d x \Bigg[ \sigma'^2 - ( \nabla \sigma)^2
+ \Big(
 \frac{(d-4) (d-2)}{4} \big( \frac{a'}{a}\big)^2
+ \frac{d-2}{2} \, \frac{a''}{a} - m^2 a^2 \Big) \sigma^2\Bigg] \, ,
\ea
in which a prime denotes the derivative with respect to the conformal time. 

Going to the Fourier space and expanding the field $\sigma$ in terms of the creation and annihilation operators we have 
\ba
\sigma(x^\mu) =  \int \frac{d^{d-1}\bfk}{(2 \pi)^{(d-1)/2}}  \Big( \sigma_k (\tau) e^{i \bfk\cdot \bfx}  a_{\bfk} +  \sigma_k (\tau)^* e^{-i \bfk\cdot \bfx} a_{\bfk}^\dagger
\Big)\, .
\ea

First we have to solve the mode function $\sigma_k (\tau)$ in a dS spacetime. 
Considering a dS spacetime with the Hubble expansion rate $H$, we have $a H \tau=-1$ where we have taken $-\infty< \tau <0$. Note that since there is no rolling scalar field (i.e. there is no inflaton field), then the dS symmetry is exact and in particular 
$\dot H=0$, yielding to $a'= H a^2$ and $a''= 2 H^2 a^3$. 
Then the equation of motion for the mode function $\sigma_\bfk(\tau)$ is given by
\ba
\label{KG-eq}
\sigma_k''  + \Big[ k^2 +  \frac{m^2}{H^2 \tau^2} - \frac{(d-2)^2}{2\tau^2}   \Big] \sigma_k =0 \, .
\ea 
As a check note that if we set $d=4$, then the last term in the bracket above 
yields the well known contribution $-2/\tau^2$ which is the hallmark of gravitational
particle production in a dS background for the light  scalar field  perturbations.
  
If we expect Eq. (\ref{cc-formula1}) to hold true, then $H \sim m^2/M_P$ and therefore $H/m \sim m/M_P \ll1$. Therefore, we can safely ignore the last term 
in Eq. (\ref{KG-eq}) compared to the mass term. Imposing the Bunch-Davies (Minkowski) vacuum for the modes deep inside the dS horizon $k \tau \rightarrow - \infty$, 
\ba
\sigma_k = \frac{1}{\sqrt{2 k}} e^{- i k \tau}\, ,
\ea
the solution of the mode function is given in terms of the Hankel function of the first type as follows
\ba
\label{mode-Hankel}
 \phi_k (t) =   a^{\frac{2-d}{2}} \sigma_k =  (- H \tau)^{\frac{d-1}{2}} \big(\frac{\pi}{4 H} \big)^{\frac{1}{2}} e^{-\frac{\pi}{2} \nu} H_{i \nu}^{(1)}(- k \tau)  \, ,
\ea
where 
\ba
\nu \equiv \sqrt{ \frac{m^2}{H^2}  -1} \simeq \frac{m}{H} \, .
\ea

Having calculated the mode function $\phi_k(\tau)$ we can calculate the vacuum energy density as follows
\ba
\label{rho-dim1}
\langle \rho \rangle = \frac{\mu^{4-d}}{2 a^2  } \int \frac{d^{d-1} \bfk}{(2 \pi)^{d-1}}  
\Big[ \,  |\phi'_k(\tau)|^2 + (k^2 + m^2 a^2) |\phi_k(\tau)|^2
\Big] \, ,
\ea
in which $\mu$ is a mass scale to properly take care of the energy dimension. 

Defining the dimensionless variable $x \equiv - k \tau$, and separating the angular and the radial contributions  of the integral, we obtain 
\ba
\label{rho-dim2}
\langle \rho \rangle = \frac{\pi \mu^{4-d}}{ 8 (2 \pi)^{d-1} } e^{-\pi \nu} H^d
\big( \int d^{d-2} \Omega\big)  \int_0^\infty x^d 
\Big[ \,  \Big | \frac{d}{dx} H_{i \nu}^{(1)} (x) \Big|^2 + (1 + \frac{\nu^2}{x^2}) \Big|H_{i \nu}^{(1)} (x) \Big|^2
\Big] \, .
\ea
In deriving the above expression, we have neglected the time derivative of the prefactor $(- H \tau)^{(d-1)/2}$ in the mode function Eq. (\ref{mode-Hankel}). One can check that this is a very good approximation in the limit of interest here where
$\nu \gg 1$. The fractional errors induced in the final result is smaller by a factor  $1/\nu^2 \ll 1$.

The integral over the azimuthal directions in Eq. (\ref{rho-dim2}) is given by  \cite{Martin:2012bt}
\ba
\int d^{d-2} \Omega = \frac{2 \pi^{(d-1)/2}}{\Gamma(\frac{d}{2} - \frac{1}{2})}  \, ,
\ea
where $\Gamma(x)$ is the gamma function. 

The final step is to perform the following integral
\ba
I \equiv \int_0^\infty x^d  \Big[ \,  \Big | \frac{d}{dx} H_{i \nu}^{(1)} (x) \Big|^2 + (1 + \frac{\nu^2}{x^2}) \Big|H_{i \nu}^{(1)} (x) \Big|^2
\Big] \, .
\ea
Fortunately the above integral can be taken exactly yielding \footnote{We use Maple computational package to perform the integral.}
\ba
\label{I-value}
I &=& \frac{(1- d + 2 i \nu) }{4 \pi^{5/2} \sinh^2(\nu \pi) }  \, 
\Gamma( -i\nu -\frac{1}{2} + \frac{d}{2}) \,  \Gamma( i\nu +\frac{1}{2} + \frac{d}{2}) \, \Gamma( \frac{d}{2} - \frac{1}{2}) \,  \Gamma( - \frac{d}{2}) \,  \times {\cal C} \, ,
\ea
where ${\cal C}$ is given by
\ba
{\cal C} &\equiv&   2 \cosh(\nu \pi) \cos(\frac{\pi ( 2 i \nu - d)}{2}  ) \cos(\frac{\pi ( 2 i \nu + d)}{2}  )  \nonumber\\
&-& \cos(\frac{\pi d}{2}) \left[\cos(\frac{\pi ( 2 i \nu + d)}{2}  ) + \cos(\frac{\pi ( 2 i \nu - d)}{2}  )  \right] \, .
\ea
The singular limit of the integral $I$ governing the UV divergence of the zero point energy is controlled by the term $\Gamma( - \frac{d}{2})$ in Eq. (\ref{I-value}).

To perform the dimensional regularization, we expand $d= 4 -\epsilon$ in $\langle \rho \rangle$, obtaining
\ba
\label{rho-reg}
\langle \rho \rangle = \frac{H^4}{2048 \pi^2} ( 4 \nu^2 + 1) (4 \nu^2 + 9) \Big[ -\frac{4}{ \epsilon} + 2\ln \left(\frac{H^2  }{4 \pi \mu^2} \right) + \Delta + {\cal O} (\epsilon )
\Big] \, ,
\ea
where
\ba
\Delta \equiv  2 \Psi(\frac{5}{2} + i \nu) +   \Psi(\frac{3}{2} - i \nu)+
\frac{(8 \gamma - 12) \nu^2 +8 i \nu  + 18 \gamma -15}{ (4 \nu^2 + 9)   } \, ,
\ea
in which $\Psi(x)$ is the Polygamma function and $\gamma$ is the Euler-Mascheroni constant. One can check that for large $\nu$ limit (which is the case here)  $\Delta \rightarrow 4  \ln(\nu)$. Finally, note that the $\frac{1}{\epsilon}$ term in Eq. (\ref{rho-reg}) originates from $\Gamma( - \frac{d}{2})$ which is divergent near $d=4$. 

Now expanding $\langle \rho \rangle$ to leading order in $\epsilon$  and taking the limit
$\nu \simeq m/H \gg 1$,  we obtain
\ba
\label{rho-dS-reg}
\langle \rho \rangle = -\frac{\nu^4 H^4}{64 \pi^2}  \left[\frac{2}{\epsilon} - 
\ln \left(\frac{H^2 \nu^2 }{4 \pi \mu^2} \right)  +...\right] \, ,
\ea
where the  terms in ... are at the order $\nu^{-2}$ and smaller which are subleading compared to the log term.   

As expected, the divergent $1/\epsilon$ term in Eq. (\ref{rho-dS-reg}) 
represents the logarithmic divergence of $\langle \rho \rangle$ with the UV scales. After regularizing  this divergent term and absorbing the factor $4 \pi$ 
 into  our renormalization scale and noting that $\nu \simeq m/H$, we recover the desired formula 
 \ba
\label{cc-formula-dS}
\langle \rho_v \rangle =    \frac{m^4}{64 \pi^2} \ln( \frac{m^2}{\mu^2}) + { \cal O} \big(m^2 H^2 \big) \, ,
\ea
in agreement with  Eq. (\ref{cc-formula1}) obtained in a flat background up 
to subleading corrections ${ \cal O} (\nu^{-2})$. 
It is intriguing that the factor $H$ drops out to leading order  both in the prefactor and inside the log term in Eq. (\ref{rho-dS-reg}).

There are a few comments in order. First, while the leading order in Eq. (\ref{cc-formula-dS}) agrees with Eq. (\ref{cc-formula1}) in flat background, but we see that there are subleading corrections as well. These subleading corrections contain various even powers of $H$, such as $m^2 H^2$ and $H^4$. This is because the dS background has a curvature radius of order  $H^{-2}$ which should enter the analysis.  This is not inconsistent with our starting argument based on equivalence principle. More specifically, the equivalence principle can pin down the local term which is independent of the curvature radius of the spacetime. However, the subleading terms containing powers of $H$ involve the curvature radius of the spacetime and as such inaccessible via equivalence principle. Second, for the massless term, the first two leading terms  vanish and we end up with the $H^4$ contribution in $\langle \rho_v \rangle$. This is again a unique feature of the curved background which is not accessible for flat spacetime. Third, as shown in \cite{Onemli:2002hr}, for  the massless field with a quartic coupling $\lambda$, the contribution $\langle \rho_v  + P_v\rangle \propto \lambda H^4$ indicating the violation of the weak energy condition. This is again a non-trivial property of the
vacuum stress energy tensor in a curved background.

To complete the discussions, we also calculate the density contrast $\delta \rho/\langle \rho \rangle$ in this background. The analysis is similar to what were performed in section \ref{scalar-contrast}. As before, let us define  
the three contributions $\rho_i$ into energy density as in Eq. (\ref{rho-i-def}) 
in which the specific form of $\langle \rho_1 \rangle, \langle \rho_2 \rangle$ 
and $\langle \rho_3 \rangle$ are given respectively in Eq. (\ref{rho-dim1}). 
In addition, performing the dimensional regularization for each component of
$\langle \rho_i \rangle$ one confirms that the relations in Eq. (\ref{rho123}) are valid here as well with  subleading corrections ${ \cal O} (\nu^{-2})$. Furthermore, performing the same steps as in section \ref{scalar-contrast}  
one can easily check that $\langle \rho_3^2 \rangle$ and $\langle \rho_2^2 \rangle$ still satisfy the relations (\ref{rho-squared1}) and  (\ref{rho-squared2}) respectively. However, there are new contributions in  $\langle \rho_1^2 \rangle$ due to the fact that $\phi(x^\mu) = \sigma(x^\mu)/a(\tau)$ and $\dot a = a H$.  
One can check that the new contributions in $\langle \rho_1^2 \rangle$ compared   to Eq.  (\ref{rho-squared1}) are suppressed  by the factor $\frac{H^2}{m^2} \simeq \nu^{-2} \ll1$, so $\langle \rho_1^2 \rangle = \langle \rho_1 \rangle^2 \big(3+ { \cal O} ( {\nu^{-2}} ) \big)$.

Correspondingly, combining all combinations in $\delta \rho$ as in 
Eq. (\ref{delta-rho}), the density contrast in dS background is obtained to be 
\ba
\label{contrtast-dS}
\frac{\delta \rho_v}{\langle \rho_v \rangle} = \pm \sqrt{10}
 + { \cal O} \big( {\nu^{-2}} \big) \, .
\ea
It is both interesting and  reassuring that the expressions for $\langle \rho_v \rangle$
and $\frac{\delta \rho_v}{\langle \rho_v \rangle}$ match the corresponding formula in the flat background up to subleading corrections ${ \cal O} \big( {\nu^{-2}} \big)$.

Similar to the case of the flat background one can also calculate the correlation length  associated with the zero point energy density fluctuations 
in the dS background. The analysis are somewhat complicated due to technicalities associated with the Hankel functions. Using the approximate relations of the Hankel functions we have verified that the correlation length of the zero point fluctuation in dS backgrounds is indeed $\xi \sim 1/m$ as in flat background. This is consistent with our notion that the intrinsic properties of the zero point energy density fluctuation such as its amplitude and the correlation length are local phenomena and are insensitive to the large scale structure of the spacetime.    
 
Now equipped with the exact mode function in a dS background, we can calculate 
the accumulated energy density on super Compton scale $\delta \rho_C$ and compare the result with its flat counterpart Eq. (\ref{rhovC}). Setting simply $d=4$ in Eq. (\ref{rho-dim2}) and $\nu \simeq m/H$, we obtain
\ba
\label{super-Compton}
\delta \rho_C &\simeq&  \frac{H^4}{16 \pi}  \int_0^\nu x^4  \Big[ \,  \Big | \frac{d}{dx} H_{i \nu}^{(1)} (x) \Big|^2 + (1 + \frac{\nu^2}{x^2}) \Big|H_{i \nu}^{(1)} (x) \Big|^2
\Big] \, ,  \nonumber\\
&\simeq& \frac{H^4}{16 \pi} \frac{\nu^4}{\pi} \simeq \frac{m^4}{16 \pi^2} \, .
\ea
Interestingly, we see that the  above value of $\delta \rho_C$ is in good agreement 
with its flat counterpart given in Eq. (\ref{rhovC}). 
  
As we have mentioned above,  the correlation length of zero point energy density is $\xi \sim 1/m$ so on length scale far beyond this correlation length the distribution of the zero point energy is expected to be uncorrelated.  
On the other hand, from Eqs. (\ref{cc-formula-dS}) and (\ref{super-Compton}) we see that the accumulated energy density on super Compton scales (i.e. for modes with wavelength larger than  $\xi$) is comparable to the total zero point energy, $\delta \rho_C/\langle \rho_v \rangle \sim 1$. This again confirms that the zero point energy of the heavy fields  can not be used to cover the entire spacetime as a background energy density. In other words, one needs to impose $\langle \rho_v \rangle \ll \rho_T \simeq \rho_F$ in order to have a sensible cosmological background. 

\section{Cosmological Implications}
\label{implications-sec}

The results from the previous three Sections  excludes the heavy fields with 
$\langle \rho_v \rangle \gg \rho_F$
 to contribute to the observed dark energy today. Originally  we came to this conclusion based on the fact that one requires an enormous number $N_{\mathrm{patches}}  \gg 1$ of incoherent small patches of size $H_{(m)}^{-1}$ to cover the FLRW horizon today. This conclusion was specifically confirmed in Section \ref{variance-sec} by looking at the density contrast $|\frac{\delta \rho_v}{\langle \rho_v \rangle}| \sim 1$.  A natural question now is  if the heavy fields do not contribute to the current dark energy (cosmological constant) then what roles in cosmology  do their vacuum energy density play?

To answer the above question we again employ the picture based on the scale of dS horizons as presented in Fig. \ref{H0-Hm}. Consider the early stage in cosmic expansion history when $\rho_F \gg \rho_v$ so 
$H_F^{-1} \ll H_{(m)}^{-1}$ as in left panel of Fig. \ref{H0-Hm}. In this situation the observable universe is inside the horizon of zero point energy but at the same time since $\rho_v \ll \rho_F$, then the contribution of $\rho_v$ in expansion rate is too small to be important. As time proceeds and $\rho_F$ decreases further  we approach the epoch when $\rho_F \sim \rho_v$ and the zero point energy becomes relevant and we notice the effects of the vacuum energy. As time goes by further, $\rho_F$ falls off rapidly so we end up with the situation $\rho_v \gg \rho_F$ as in right panel of Fig. \ref{H0-Hm}.  In this case the regions 
filled with the zero point energy develop strong inhomogeneities while falling into the FLRW Hubble horizon. The time scale for the dS patches to create inhomogeneities 
 is about $2/H_{(m)}$ as it takes time of about $1/H_{(m)}$ for each dS horizon to enter the FLRW horizon.  Once the FLRW horizon grows large enough (or equivalently when $\rho_F \ll  \rho_v$ ) more and more of dS patches enter the FLRW Hubble radius. The variance condition (\ref{fraction}) indicates that the mass inside the dS patches inside the FLRW Hubble radius may collapse to form black holes. However, as we mentioned at the end of section \ref{scalar-contrast}, the formation of black hole from the collapse of local AdS regions created from inhomogeneities is far from obvious  as the interplay between the turbulent dynamics of local AdS and dS inhomogeneities are complicated.  But if the 
masses inside these patches collapse into black holes, their contribution in background cosmological dynamics is expected to be in the form of matter. This indicates that the resulting mass may behave like dark matter or the seeds of dark matter.

Despite the above qualitative discussions, at this stage we can not be 
more specific about the subsequent contribution of the zero point energy of heavy fields after the time when  $H_F^{-1} > H_{(m)}^{-1}$. Different scenarios are possible which we study below using a phenomenological  fluid description.

\subsection{Phenomenological Fluid Description}

As mentioned above we do not know the detail mechanism which governs the 
contribution of the zero point energy of heavy fields after the time when  $H_F^{-1} > H_{(m)}^{-1}$. Here,  we consider a phenomenological approach and treat the effects  of heavy fields like a fluid with an undetermined equation of state. 

To be more specific, let us denote the energy density of massive fields after the time when $\rho_F$ is diluted enough and 
$\rho_F < \rho_m$ by $\tilde {\rho}_m$. We also assume that each components 
of effective energy density to evolve separately by their own energy conservation:
$\dot \rho_F + 3 H (\rho_F + p_F) =0$ and $\dot{\tilde\rho}_m + 3 H (\tilde\rho_m + \tilde p_m) =0$ in which $\tilde{p}_m$ represents the effective pressure of the new matter-like energy component.  Defining the equation of states  via $p_F \equiv w_F \rho_F$ and $\tilde p_m \equiv w_m \tilde\rho_m$, the total energy density now is $\rho_T = \rho_F + \tilde\rho_m$, yielding
\ba
\rho_T(t) = \Big[ \rho_m \Big( \frac{a(t)}{a(t_m)}\Big)^{-3 (1+ w_m)} + 
\rho_F(t_m) \Big( \frac{a(t)}{a(t_m)}\Big)^{-3 (1+ w_F)}
\Big] \, ,
\ea
in which $t_m$ is the time when $\rho_F(t_m) \sim \rho_m $, i.e. when the two Hubble radius $H_F^{-1}$ and $\Hm^{-1}$ nearly coincide. Considering the numerical uncertainties, we set $ \tilde\rho_m(t_m) \equiv \kappa' \rho_F(t_m)$ in which 
$\kappa'$ is a numerical constant which may depend on $m$. Then, the total energy density is given by 
\ba
\label{rho-m-tot}
\rho_T(t) = \rho_F(t_m)  \Big( \frac{a(t)}{a(t_m)}\Big)^{-3 (1+ w_F)}
\Big[ 1+  \kappa' \Big( \frac{a(t)}{a(t_m)}\Big)^{3 (w_F - w_m)} 
\Big] \, .
\ea

The above is a phenomenological description of the contribution of the heavy fields after the time when $\rho_F < \rho_m$. We have introduced the effective equation of state parameter $\omega_m$ to capture the uncertain behaviour of the zero point energy associated with the massive fields. If strong inhomogeneities with $\delta \rho_v/\rho_v \sim 1$ yields to  black hole formation then  $\omega_m =0$ and $\tilde\rho_m$ can play the roles of dark matter. Of course this is a good news in which the  nature of dark matter and dark energy is unified in this proposal as both being generated dynamically from the zero point energy. However, the difficulty with $\omega_m \simeq 0$ is that at the early time in cosmic history,
the produced dark matter from the zero point energy of electron and heavier fields 
rapidly dominate over radiation energy density, long before the time of matter radiation equality at the temperature $T_{eq} \sim 3\,  \mathrm{eV}$,
altering the hot big bang cosmology in various unwanted ways. One way out may be to consider a situation in which the phenomenological parameter $\kappa'$ in Eq. (\ref{rho-m-tot})  is  (exponentially) small  so it will take a long time for the seed of dark matter to take over the radiation energy density.
On the other hand, if $\omega_m \simeq \frac{1}{3}$ then $\tilde\rho_m$ may behave like dark radiation.  Considering the uncertainties involved in the process, $\omega_m$ may take different values for various  fields  so $\tilde\rho_m$ may behave differently during cosmic history so we may have dark radiation, dark matter or stiff fluid with $\omega_m \lesssim 1$.

Another possibility is that the effective fluid associated  to the zero point energy  is such that $\omega_m = \omega_F$ so $\tilde\rho_m(t) \propto \rho_F(t)$, 
i.e.
\ba 
\label{rho-heavy}
\tilde\rho_{(m)}(t) = 3\kappa M_P^2 H_F^2 = \kappa \rho_F \, ,
\ea
in which $\kappa$ is a numerical constant.  
This corresponds to a tracking scenario 
 in which  the energy density of the 
heavy fields at each stage in cosmic history tracks the background FLRW energy density. During the radiation dominated era their energy densities follow that of a radiation energy density while during the matter dominated era they behave like matter. This is an interesting case so below we  concentrate on this scenario more closely.    
 
 The tracking scenario has interesting implications for the current energy density of dark matter. To see this, let us  assume that $\rho_F= \rho_\gamma +\rho_b$ in which $\rho_\gamma$ and $\rho_b$ represent the energy density of radiation and baryon. Specifically, we assume that there is no conventional (say WIMPs-like) dark matter in our setup. Let us define the time of matter-radiation equality at the redshift $z_{eq}$ as when $\rho_b(z_{eq}) = \rho_\gamma(z_{eq})$. Now observe that what we mean by dark matter in $\Lambda$CDM model is the zero point energy density given by the heavy field in Eq. (\ref{rho-heavy}). Specifically,  the current would-be dark matter energy density is actually $\rho_{DM}^{(0)}= \tilde\rho_{m}(t_0) =
\kappa \rho_F^{(0)}=  
\kappa \rho_b^{(0)} + \kappa \rho_{\gamma}^{(0)}$. 
Taking into account that the energy density of radiation and baryon falls off  like $a^{-4}$ and $a^{-3}$ respectively we obtain 
\ba
1+ z_{eq} = \frac{1}{1+ \kappa} 
\Big(  \frac{\Omega_M^{(0)}}{\Omega_\gamma^{(0)}} -\kappa \Big) \, ,
\ea
in which $\Omega_M^{(0)} \sim 0.31$ $(\Omega_\gamma^{(0)} \sim 10^{-4}) $
is the fractional total matter (radiation ) energy density today
as inferred from $\Lambda$CDM model. Now defining $z_{eq}^{\Lambda CDM}$ as the time of matter-radiation equality in $\Lambda$CDM model and assuming $\kappa \sim  {\cal O} (1) $
we obtain
\ba
z_{eq} \simeq \frac{z_{eq}^{\Lambda CDM}}{1+\kappa} \, .
\ea
This means that in the current setup the time of matter-radiation equality is shifted towards later epoch in cosmic history. This is because we have assumed that the time of matter-radiation equality happens when $\rho_b(z_{eq}) = \rho_\gamma(z_{eq})$.
Taking $\kappa \lesssim 1$ we may have $z_{eq} \sim  \mathrm{few} \times 1000$. 
However, note that shifting $z_{eq}$ close to the time of CMB last scattering can be dangerous for the CMB observations. But we also note that shifting the time of matter-radiation equality towards later epoch can play important roles   to solve  the $H_0$ tension problem.

The above conclusion about the roles of the heavy fields may seem in conflict with our starting point that the vacuum energy should be locally Lorentz invariant and
$\langle p \rangle = - \langle \rho \rangle$ which is trivially violated  for both radiation  
and matter. The answer is that the relation $\langle p \rangle = - \langle \rho \rangle$ is enforced locally, say deep inside each dS patch. However, what we  have  in Eqs.   (\ref{rho-heavy}) or (\ref{rho-m-tot}) is the collective energy density of a large number of patches  of zero point energy inside the FLRW horizon 
which may not be in conflict with the local 
requirement $\langle p \rangle = - \langle \rho \rangle$ deep inside each dS patch. 


\subsection{Selection Rules}

Based on the results from the previous Sections, we end up with  an interesting ``selection rule" which works as follows. At each stage in cosmic epoch, only  field 
with $\Hm \sim H_F$ and energy density   $ \langle \rho_v \rangle \sim \rho_F $  can be relevant as the source of  the  dark energy.  
Fields which are much lighter ($\Hm \ll H_F$) are irrelevant in cosmic expansion.  This is simply because their contributions in $\langle \rho_v \rangle$ as given in Eq. (\ref{cc-formula1})  is exceedingly smaller than $\rho_F$ to be noticeable. Finally,  heavy fields with $\Hm \gg H_F$ can not be the source of dark energy while they may form the seeds of dark matter from the start or to track the background FLRW energy density $\rho_F$ and be the
source of dark matter after the time of matter-radiation equality.  Alternatively, they may behave like a stiff fluid with $\omega_m \lesssim 1$ in which their contributions in background energy density is diluted faster than radiation after the time $t> \Hm^{-1}$.

The above selection rule can easily address both the old and the new cosmological constant problems. Recall that the old cosmological constant problem is why $\rho_v$ is not  large i.e. why it is so small as $(10^{-3} \mathrm{eV})^4$. The new cosmological constant problem is why the vacuum energy density becomes comparable to the matter energy density at the current stage of the cosmic history. The resolution is that there is a field in the SM field content, the lightest neutrino{\footnote{While the differences in neutrino mass squared are known, but the absolute masses and the mass of lightest neutrino are not exactly known, see \cite{Zyla:2020zbs}. Here we adopt the simple assumption that the lightest neutrino has the mass at the order $10^{-2} \mathrm{eV}$ \cite{Loureiro:2018pdz, GAMBITCosmologyWorkgroup:2020rmf}.}},  with the vacuum zero point energy 
$\rho_{(\nu)} \sim m_\nu^4$  which happens to have a mass at the same order as $\rho_{F 0}^{1/4} \simeq  \rho_c^{1/4}$
in which $\rho_c \sim (10^{-3} \mathrm{eV})^4$ is the critical energy density.  The entire FLRW universe is currently within a single patch of the lightest neutrino with the horizon radius $H_{(m_\nu)}^{-1}$. The current  dark energy survives in future for another period of roughly $1/H_{(m_\nu)} \sim 1/H_0 \sim 10^{10}$ years before multiple dS patches of the lightest neutrino enter the FLRW horizon.  Of course, this conclusion about the future dynamics of the universe is based on the assumption that there is no light field in beyond SM sector (say like light axion) with mass lighter than 
$10^{-3} \mathrm{eV}$ to contribute to the future dark energy. 
We comment that the relations between the zero point  energy of neutrino and the observed dark energy (cosmological constant) have been studied in different contexts in  the past  in \cite{Shapiro:2000dz, Shapiro:1999zt, Glavan:2015ora, Blasone:2004hr, Ibanez:2017kvh}.

An immediate corollary of our selection rule is that at early stage in cosmic history (i.e. for all time prior to the time of matter-radiation equality)
only fields with the masses at the order of the FLRW photon temperature could contribute to dark energy at that epoch. This is because the energy density of the FLRW background during the radiation dominated era 
is related to the photon temperature via
\ba
\rho_F = \frac{\pi^2}{30} g_* T^4 \, ,
\ea
in which $g_*$ is the effective relativistic degrees of freedom at the temperature $T$.   Now  comparing the above equation with Eq. (\ref{cc-formula1})  for the vacuum energy density, and noting that for the field of interest $\Hm \sim H_F$,  
we conclude that
\ba
\label{m-T}
m \sim T.
\ea
If a field is much lighter than $T$, then its vacuum energy density $\propto m^4$ 
is much smaller than $T^4$ and can not be important in cosmic energy density. On the other hand, if a field is much heavier than $T$, then based on the arguments mentioned  in previous Sections the dS horizon radius associated with  its vacuum energy density is much smaller than the Hubble radius $H_F^{-1}$  and they collapse into FLRW horizon behaving like dark radiation, dark matter or like a stiff fluid.  Note that while formula (\ref{m-T}) works well for the time prior to matter-radiation equality, but it also gives a reasonable estimation for the relation between the mass of the lightest neutrino and  the energy scale of the current dark energy. To see this note that $T_{\gamma_0} \sim 10^{-4} \mathrm{eV}$ which is only about two orders of magnitude below the expected mass
of lightest neutrino. Of course, this is easily understandable since after the time of matter-radiation equality the energy density of radiation falls off by an additional factor  of $1/a$ compared to the matter energy density.  This is why we had to use the critical mass density $\rho_c$ instead  of $T_{\gamma0}^4$ to correctly estimate the magnitude of the current dark energy density.  

 Since only fields with $m \neq 0$ contribute to the vacuum energy we conclude that  before the electroweak symmetry breaking at the temperature 
 $T_{EW} \sim 160 \,  \mathrm{GeV} $,  all SM fields were massless and they did not contribute to dark energy.  Immediately after the electroweak phase transition the relevant fields are the top quark $m_t \simeq 170 \,  \mathrm{GeV}$, the Higgs field $m_{H} \sim 125 \,  \mathrm{GeV}$ 
 and the three gauge bosons $m_Z \simeq 91 \,  \mathrm{GeV}, m_{W^{\pm}} \simeq 80 \, \mathrm{GeV}$. The top quark and the Higgs field are somewhat heavy so they are at the threshold of being able to contribute into dark energy at that time, while the three gauge bosons have good chances to play important roles as the source of dark energy. Of course, as the temperature of the  universe falls below the mass of vector bosons, their contribution in dark energy expires and they may contribute to dark radiation, dark matter or behaving like a stiff fluid.

This story is repeated when the background temperature approaches the mass of other fundamental particles such as $m_{\tau} \simeq  1.7 \,  \mathrm{GeV} $,
$m_{\mu} \simeq  105\,  \mathrm{MeV}$ and $m_{e} \simeq  0.5 \,  \mathrm{MeV} $. Specially important is the time when the temperature is around $\mathrm{MeV} $ when the big bang nucleosynthesis (BBN) is at work. Changing the Hubble expansion rate of the universe at this epoch, either by contribution from dark energy of the electron field or its contributions in the form of  dark radiation, 
can affect the  dynamics of  BBN. This issue requires careful investigation which can provide a non-trivial check for the consistency of the 
whole picture presented here.   Another important time to look for is when the heavier neutrino fields (i.e. not the lightest neutrino) with the mass $m \sim 0.1 \,  \mathrm{eV}$ become relevant to contribute into dark energy. This is somewhat around $T \sim 10^{-2} \,   \mathrm{eV} $ and redshifts of $z \sim 10^2$. 

Based from the above discussions, we have transient periods of dark energy  for a few e-folds anytime in cosmic history when the condition $m \sim T$ is met during which the Hubble expansion rate stays nearly constant and then it falls off as in conventional big bang cosmology. These can happen in cosmic history both after the surface of last scattering (for heavier neutrinos) at redshift $z \sim 10^2$  and before the surface of last scattering (for $e$, $\mu$ and $\tau$ fields) at much higher redshifts.   The curious conclusion is that since the energy density is nearly constant for  a few e-folds then the value of the Hubble expansion rate today (and at the time of last scattering) 
would be larger than what the standard $\Lambda$CDM model predicts. Multiple transient periods of dark energy both at early and intermediate cosmic expansion history is a non-trivial prediction of the model that has good potentials to solve the $H_0$ tension problem \cite{Riess:2019cxk, Riess:2021jrx, DiValentino:2021izs, Dainotti:2021pqg, Dainotti:2022bzg}.  In this view, our proposal can incorporate the early dark energy (EDE) mechanism \cite{Poulin:2018cxd, Smith:2019ihp} which is proposed to  solve the $H_0$ tension problem.


\section{Summary and Discussions}
\label{sum-sec}

In this work we have revisited the quantum  cosmological constant problem.
As already known in the literature,  the conventional approach of imposing a hard UV momentum cutoff violates the underlying Lorentz invariance of the vacuum. 
Employing a regularization scheme which respects the underlying Lorentz invariance of the theory, such as the dimensional regularization method, one obtains that the energy density of each field scales like $\rho_v \sim m^4$. Furthermore, the condition of Lorentz invariance of the vacuum, $\langle p \rangle = - \langle \rho \rangle$ is manifest.   Some noticeable conclusions are that the massless fields such as gravitons, photons and gluons do not contribute to the vacuum energy density. Furthermore, $\langle \rho\rangle$  runs logarithmically with the renormalization scale $\mu$ and depending on the ratio  $m/\mu$ the value of cosmological constant can be either positive (dS) or negative (AdS). 

We have highlighted that the dS horizon associated with  the zero point energy plays important roles. For a  field with mass $m$ 
the associated dS space has a horizon radius $\Hm^{-1} \sim M_P/m^2$ which was not taken into account in previous studies of the cosmological constant problems. In the conventional approach, it is naively assumed that the entire observable FLRW patch is covered by the vacuum energy density $\langle \rho\rangle  \sim m^4$ irrespective of the vast hierarchy between the two Hubble radii 
$H_F^{-1}$ and $\Hm^{-1}$. For the heavy fields with the large hierarchy 
$\Hm \gg H_F $ this requires as many as $ \big( \frac{H_{(m)}}{H_F}\big)^3 \gg 1$ independent patches of zero point energy to fill the FLRW patch. However, since the patches of zero point energy are created quantum mechanically they are uncorrelated. Therefore, one expects that the space filled with as many as $ \big( \frac{H_{(m)}}{H_F}\big)^3 $ uncorrelated patches of  zero point energy to be highly chaotic and may even collapse to form  black holes. To support this picture, we have calculated the variance of the zero point energy for both the real scalar field and the Dirac fermion field. In both cases we have found  that $|\delta \rho_v/\langle \rho_v \rangle| \sim 1$ supporting the above intuitive picture. Furthermore,  the correlation length of the fluctuations of the zero point energy is calculated to be at the order 
$\xi \sim m^{-1}$. This results in the vast hierarchy $\xi \ll \Hm^{-1} \ll H_F^{-1}$. It is shown that the accumulated energy density on super Compton scales, i.e. 
scales which are far beyond the correlation length, is comparable to the would-be background vacuum energy density. This conclusion, in conjunction  with the above hierarchies amongst the three scales $\xi, \Hm^{-1} $ and $H_F^{-1}$, is another sign  that one can not use the vacuum energy density $\langle \rho_v \rangle $ as the dominant background energy density. To have a viable cosmological background, we have to impose 
$\langle \rho_v \rangle  \lesssim \rho_F$ at each stage in cosmic expansion history in which 
$\rho_F$ is the non-vacuum energy density (i.e. classical energy density) of the background FLRW cosmology.

We have established a ``selection rule'' stating that at each stage in cosmic expansion history, only fields with mass $m$  satisfying the condition $\Hm \sim H_F$ can contribute to dark energy at that epoch. Specifically, only fields with the mass at the order of the background photon temperature, $m\sim T$, are relevant 
as the source for dark energy.   Lighter fields carry so little energy ($m^4 \ll T^4$) to 
be important in cosmic energy budget. On the other hand,  heavy fields with $\Hm \gg H_F$ (or $m \gg T$) can not be the source of dark energy  while they  can be the seeds of dark matter, dark radiation or a stiff fluid with equation of state 
$\frac{1}{3} < \omega_m \leq1$.    

We have speculated that both the old and new cosmological constant problems can be addressed readily.  Specifically,  these puzzled  are solved by noting that there is a field in SM spectrum, the lightest neutrino, which happens to have a mass at the order of $\rho_c^{1/4} \sim 10^{-2}\mathrm{eV}$. Another important implication is that at various stages in cosmic history the energy density experiences loitering stages of dark energy in which the 
Hubble expansion rate stays nearly constant for about one or two e-folds in expansion rate and then the energy density falls off as in $\Lambda$CDM model. As a consequence, the Hubble expansion rate at the time of CMB last scattering is expected to be higher than what is inferred from $\Lambda$CDM model. Currently we are at the last transient stage associated with the lightest neutrino. However, prior to this stage there were two transient  phases of dark energy happening at around the mass scale of the heavier neutrinos at $T \sim 10^{-2} \mathrm{eV}$ corresponding to  the redshift $z \sim 10^2$. This period will be  sometime after the time of CMB decoupling. Multiple transient phases of dark energy, both before and after the surface of last scattering, yield to a higher values of the current Hubble expansion rate compared to what is inferred from the $\Lambda$CDM model. In this view our proposal can incorporate the  EDE proposal 
 and has the potential  to solve the $H_0$ tension problem. Another interesting prediction of the model (as mentioned above) is that there may be no dark matter. All matter are in the form of known baryonic matter while the roles of  dark matter are played by the heavy fields which may track the background FLRW energy density. As such, the time of matter-radiation equality is shifted towards later time in cosmic history. 

There are a number of open questions in this study. The  immediate one is how the collapse for the patches of  zero point energy associated to  heavy fields happen. We provided only a rough phenomenological picture of this mechanism. This is an important question which controls whether or not we have the tracking energy or a fluid with the equation of state $\omega_m \sim 0$. Another open  question  
is what is the natural value of the renormalization scale $\mu$? Note that the vacuum energy runs logarithmically with $\mu$. Furthermore, depending on the statistics of the field (i.e. being a boson or a fermion) and on the ratio $m/\mu$, the vacuum energy can be either positive (dS) or negative (AdS). To solve the cosmological constant problems we have implicitly assumed that the resulting vacuum energy is positive so taking the neutrinos as the relevant fields for this purpose, we require   
$\mu >  \mathrm {eV}$. If we take $\mu$ to be the scale of electroweak symmetry breaking, say  $\mu \gtrsim10^{2} \mathrm{GeV}$,  then all fermionic (bosonic) fields add positive (negative) contribution to the vacuum energy.  
If the energy density is AdS-type, then there maybe a fall off period 
(instead of loitering phase) in Hubble expansion rate at the corresponding stage in cosmic history. This will modify the inferred value of the Hubble expansion rate at the time of CMB last scattering. Therefore, the dS and AdS contributions for all SM degrees of freedom from the mass scale $10^{2} \mathrm{GeV}$ after the electroweak symmetry breaking down to the neutrino scale should be carefully included to see how dark energy and dark matter behave at each stage in cosmic history.

 Finally, it is important to examine the implications of the scenario for other aspects of early universe cosmology. Specifically, the interplay between this proposal and cosmic inflation is an open and interesting question. However, we comment  that our proposal is not in conflict with an early stage of inflation. More specifically, during the slow-roll inflation driven by an inflaton field of mass $m$ and the potential $V$, the dominant energy is given by the inflaton classical potential $V_c$ while the vacuum zero point energy is at the order $m^4$. Since for a rolling inflaton $V_c \simeq 3 M_P^2 H^2 \gg m^4$,  the induced energy density  from the zero point fluctuations during inflation
 is negligible. Therefore,  the quantum cosmological constant is not an issue during inflation and a large enough classical vacuum energy during inflation is stable against quantum zero point corrections.   
  
 \vspace{0.7cm}

{\bf Note added:}
After the completion of this work we have became aware\footnote{We are grateful to Jerome Martin for bringing to our attention the papers by Unruh and his collaborators \cite{Wang:2017oiy, Cree:2018mcx, Wang:2019mbh, Wang:2019mee}. } of a series of papers  
 in \cite{Wang:2017oiy, Cree:2018mcx, Wang:2019mbh, Wang:2019mee} who questioned the assumption of the homogeneity of the spacetime in the presence of the vacuum zero point energy. It was concluded, among other things, that a uniform cosmological constant can not cover the large scale spacetime and  the local spacetime is very inhomogeneous as in Wheeler's spacetime foam. Using the semiclassical GR, they have studied the cosmological implications of their proposal such as in resolving the old cosmological constant problem.  Our analysis in Section \ref{variance-sec}  for the fluctuations in the zero point energy density  and large density contrast is conceptually in line with their investigation.

\vspace{1cm}

{\bf Acknowledgments} 
\vspace{0.5cm}

We would like to thank Xingang Chen, Jerome Martin, 
Shinji Mukohyama,  Mohammad Hossein Namjoo and 
Misao Sasaki for insightful comments about the draft. 
We also thank Hossein Moshafi for the preliminary investigations of the predictions
of the proposal for the $H_0$ tension problem and Omid Sameie for useful discussions.

\end{document}